\documentclass[USenglish]{article}	
\usepackage[utf8]{inputenc}				%(only for the pdftex engine)
\usepackage[big,online]{dgruyter}	%values: small,big | online,print,work
\usepackage{lmodern} 
\usepackage{microtype}
\usepackage{bbm}
\usepackage{subfigure}
\usepackage{natbib}
\usepackage{enumitem}
\usepackage{comment}
\usepackage[dvipsnames]{xcolor}
\usepackage{cancel}
\usepackage{bm}

\usepackage[normalem]{ulem} %\sout \uline
\theoremstyle{dgthm}

\theoremstyle{dgdef}

\begin{document}

%%%--------------------------------------------%%%
	\articletype{Research Article}
	\received{Month	DD, YYYY}
	\revised{Month	DD, YYYY}
  \accepted{Month	DD, YYYY}
  \journalname{De~Gruyter~Journal}
  \journalyear{YYYY}
  \journalvolume{XX}
  \journalissue{X}
  \startpage{1}
  \aop
  \DOI{10.1515/sample-YYYY-XXXX}
%%%--------------------------------------------%%%

%\title{A Bayesian Bivariate Conway-Maxwell-Poisson Model for Soccer and Baseball Scores}
%\title{A Bayesian Bivariate Conway-Maxwell-Poisson Model for Home and Away Scores in Sports Games}
%\runningtitle{Bayesian Bivariate Conway-Maxwell-Poisson Regression Model for Sports Data}

\title{Bayesian Bivariate Conway-Maxwell-Poisson Regression Model for Correlated Count Data in Sports}

\runningtitle{Conway-Maxwell-Poisson Modeling of Correlated Data in Sports}

%\subtitle{Insert subtitle if needed}

\author*[1]{Mauro Florez}
%\ use * to mark the author as the corresponding author
\author[2]{Michele Guindani}
\author[1]{Marina Vannucci} 
\runningauthor{M. Florez et al.}
\affil[1]{\protect\raggedright 
Rice University, Statistics Department, Houston, TX, e-mail: mflorez@rice.edu, marina@rice.edu}
\affil[2]{\protect\raggedright 
UCLA, Department of Biostatistics, Los Angeles, CA, e-mail: mguindani@g.ucla.edu}
	
%\communicated{...}
%\dedication{...}
	
\abstract{
	Count data play a crucial role in sports analytics, providing valuable insights into various aspects of the game. Models that accurately capture the characteristics of count data are essential for making reliable inferences. In this paper, we propose the use of the Conway-Maxwell-Poisson (CMP) model for analyzing count data in sports. The CMP model offers flexibility in modeling data with different levels of dispersion. Here we consider a bivariate CMP model that models the potential correlation between home and away scores by incorporating a random effect specification. We illustrate the advantages of the CMP model through simulations. We then analyze data from baseball and soccer games before, during, and after the COVID-19 pandemic. 
	The performance of our proposed CMP model matches or outperforms standard Poisson and Negative Binomial models, providing a good fit and an accurate estimation of the observed effects in count data with any level of dispersion. The results highlight the robustness and flexibility of the CMP model in analyzing count data in sports, making it a suitable default choice for modeling a diverse range of count data types in sports, where the data dispersion may vary.

	%The results demonstrate the robustness and flexibility of the CMP model in analyzing count data in sports, making it a suitable default alternative to model a wide range of count data types in sports, where the dispersion of the data may vary. 
	
	%We illustrate the advantages of the CMP model through simulations and the analysis of the home advantage (HA) in baseball and soccer games during the COVID-19 pandemic. 
	
	%In this paper, a bivariate Bayesian Conway-Maxwell-Poisson model is proposed to model scores in soccer and baseball. The model introduces a set of random effects to allow taking into account the correlation between the scores of the home and away teams. In simulation studies, we found that on large sample size, our model outperforms the Poisson and Negative Binomial models in terms of fitting count data, regardless of whether the data is over-, under, or equi-dispersed. Additionally, we applied the model to real data from the English Premier League of Soccer and the Major League of Baseball (MLB), analyzing the performance of the teams and the change in the Home Advantage effect on the seasons played before, during, and after the COVID-19 restrictions. Our results suggest that our model provides a flexible and robust approach to fit data with any kind of dispersion.
	}


\maketitle 
%What do we wanna say in the introduction?
%What's the literature out there on models for count data and CMP?
%What's the novelty of what we propose? Start by making bullet points, to expand later.

\section{Introduction} 

%\item Statistics in sports and the specific case of predicting soccer and baseball scores

Count data play a ubiquitous role in sports analytics, providing valuable insights across various aspects of the game. They are used to record the frequency of events or occurrences, including the number of goals scored in a soccer game, the number of strikes or balls thrown in baseball, the number of points scored in basketball, or the number of strokes taken in golf. This type of data is essential for calculating key statistics such as batting averages, shooting percentages, or save percentages, as well as to
%. Moreover, it is also commonly used in sports analytics to 
analyze and compare players' performance, team performance, and overall game strategy. For example, count data can be used to evaluate a player's ability in scoring goals or making assists, or to assess a team's defensive tactics in preventing their opponents from scoring. Count data can also be used to develop predictive models that help coaches and managers make informed decisions about game strategy and player selection. For instance, count data may be used to determine which players are most likely to score in a given situation or which plays are most likely to lead to a successful outcome. 

When analyzing such types of data, it becomes crucial to utilize models that accurately capture the distinctive characteristics of the data in order to make accurate and reliable inferences. Conventional models used to describe count data include the Poisson and the Negative Binomial distributions. For example, \cite{maher1982}, \cite{dixon1997modelling}, \cite{lee1997}, \cite{karlis2000modelling} were among the first to model the number of goals scored by opposing teams in soccer using bivariate Poisson likelihoods. Since the two teams interact during the game, the game outcomes might be correlated. Hence, models for correlated count data have also been proposed \citep{karlis2003, mchale2011modelling}, as well models using a Poisson difference distribution to model the differences in goals \cite{karlis2009bayesian} and hierarchical Bayesian models \citep{baio2010bayesian}.  The Negative Binomial distribution was originally proposed by \cite{reep1971skill} to account for the overdispersion commonly observed in baseball scores, and it has since found applications in various other sports.  However, whether to employ the Poisson or Negative Binomial distribution depends on the specific characteristics of the dataset being analyzed.  For example, 
\cite{karlis2000modelling} found no significant difference between the Poisson and the Negative binomial distributions when modeling soccer data since there is relatively little over-dispersion. In some cases, the data may even exhibit some degree of under-dispersion, as demonstrated in the analysis of data from the UEFA EURO 2020 by  \cite{fedrizzi2022uefa} and the English Premier League seasons 2015-16 to 2017-18 by  \cite{Backlund2018}.

In this paper, we propose the use of a Conway-Maxwell-Poisson (CMP) model for the analysis of count data in sports, focusing in particular on the analysis of outcomes in baseball and soccer games. The CMP distribution extends the traditional Poisson distribution by introducing an additional \emph{dispersion} parameter -- in addition to the mean -- that  % The dispersion parameter 
controls the degree of over- or under-dispersion. %, with values less than one indicating under-dispersion and values greater than one indicating over-dispersion. 
Some advantages of the CMP distribution for modeling count data, over other distributions, such as the Negative Binomial distribution, are that it is easier to interpret and more flexible in modeling data. 
Here, we illustrate these advantages in sports analytics through a series of simulations and an analysis of the home advantage (HA) in baseball and soccer games, pre-, during, and post-COVID pandemic. \cite{karlis2003} utilize a Poisson model to assess the HA effect, assuming equidispersion. In order to relax this assumption,   \cite{boshnakov2017bivariate}  propose the use of a Weibull distribution and \cite{higgs2021bayesian} employ the negative binomial distribution in the context of North American sports. However,  when there is minimal or no overdispersion present, the goodness of fit is either slightly inferior or comparable to that of the Poisson distribution.
In our modeling framework we introduce a bivariate CMP specification using a set of random effects to model within-game correlation. 
%(\cite{chib2001}).  
The approach is robust and capable of accurately fitting the data, irrespective of whether it exhibits overdispersion, equidispersion, or underdispersion. It is worth noting that \cite{piancastelli2021multivariate} also employ a multivariate Conway-Maxwell-Poisson distribution; however, their parametrization may not lend itself to a straightforward interpretation of the parameters. In addition, in our model formulation we also investigate whether the HA changed due to the COVID-19 pandemic, when crowds were prohibited or limited. Studies have shown that the HA effect was significantly reduced in soccer during this period  \citep{mccarrick2020home, tilp2020covid, benz2021, piancastelli2021multivariate}. However, a study by \cite{higgs2021bayesian} found that there was little to no change in the HA in baseball during the COVID-19 season.  Our findings also indicate a relatively smaller impact of the COVID-19 pandemic on the HA in baseball compared to soccer.

The paper is structured as follows: in Section \ref{model}, we briefly describe the Conway-Maxwell-Poisson distribution, our model, and the parametrization used in the paper. In Section \ref{inference}, we describe how we obtain posterior inference, including the MCMC algorithm. In Section \ref{simulations}, we present a simulation study where we evaluate the performance of the model and the estimation of the HA effect. In Sections \ref{data_analysis_soccer} and \ref{data_analysis_baseball}, we present our data analysis on Soccer English Premier-League and Major League of Baseball games, respectively, in seasons 2019-2022. In Section \ref{conclusion}, we provide some concluding remarks.

\section{Methods}
\subsection{A Bayesian Conway-Maxwell-Poisson regression model for sports data} 
\label{model}

We propose to model the number of points or the goals scored by the home and away teams in each game as Conway-Maxwell-Poisson (CMP or COM-Poisson) random variables. The CMP distribution, first introduced by \cite{conway1962}, is a two-parameter generalization of the Poisson distribution that allows modeling count data with over-, under-, or equi-dispersion. In this paper, we consider the reparametrization proposed by \cite{guikema2008} where the probability mass function (pmf) is expressed as
\begin{equation}
	P(Y = y \mid \mu, \nu) =  \left(\frac{\mu^y}{y!}\right)^\nu \frac{1}{Z(\mu,\nu)} ,\qquad  \mu > 0, \nu \geq 0,
\end{equation}

\noindent with $q(y \mid \mu, \nu) = \left(\frac{\mu^y}{y!}\right)^\nu$ indicating the unnormalized component of the pmf and $Z(\mu, \nu) = \sum_{y=0}^{\infty}q(y | \mu, \nu)$ the intractable normalising constant. This parametrization relies on a centering parameter $\mu$, which facilitates straightforward interpretation when fitting a generalized linear model (GLM) and a shape parameter $\nu$, which enables the modeling of various dispersion scenarios across measurements. Specifically, $\nu < 1$ accounts for overdispersion, $\nu > 1$ accounts for underdispersion, and $\nu = 1$ corresponds to equidispersion, representing the Poisson distribution. The mode of the CMP distribution is given by the flooring function, $\lfloor \mu \rfloor$, which rounds down the value of $\mu$ to the nearest integer. However, the moments of the distribution cannot be directly computed due to the analytically intractable normalizing constant $Z$. Instead, they can be approximated using asymptotic results. The expected value of the distribution can be approximated as $E(Y) \approx \mu + \frac{1}{2\nu} - \frac{1}{2}$, while the variance can be approximated as $Var(Y) \approx \frac{\mu}{\nu}$  \citep{shmueli2005}. These approximations are quite accurate, except for small $\mu$ or $\nu$ in $E(Y)$, and for $\nu > 1$ in the $Var(Y)$ \citep{benson2021}.

Let $\bm{y_{i}} = (y_{i1}, y_{i2})$ denote the observed number of goals or points in game $i$, $i=1, \ldots, n$, for the home team $(j = 1)$ and the away team $(j = 2)$. We propose a CMP generalized linear model framework to predict $y_{ij}$, where, similarly as in \cite{karlis2003}, we include a bivariate game-specific random effect specification to adjust for a team's relative strength offense, the opponent's relative strength defense, as well as to capture a Covid-19 effect and the correlation between the home and away scores. More specifically, we model the $\mu$ and $\nu$ parameters of the CMP distribution as a function of the available covariates, say $\bm{x_{ij}}$, and the latent correlated random effects:
\begin{equation*}
    y_{ij} | \mu_{ij}, \nu_{ij} \sim CMP(\mu_{ij}, \nu_{ij}), \qquad i=1, \ldots, n, j=1,2,
    %\label{model}
\end{equation*}
where the centering vector $\bm\mu=(\mu_{i1}, \mu_{i2})^T$ represents the expected number of goals or points scored by the home team and the away team playing in game $i$, respectively, and it is specified as 
%, as suggested by \cite{guikema2008}.
%More specifically, let indicate the covariates and $\mathbf{b_i} = (b_{i1}, b_{i2})$ the set of game-specific random effects. Given the values of the random effects $\mathbf{b_i}$, that captures the correlation between the goals of the home team and the away team, as shown in equation \ref{proof}, and parameters $\beta_j$, and $\gamma_j$, the number of goals or points of the home and away teams are modeled as independent COM-Poisson
\begin{gather*}
    log(\mu_{i1}) %= \bm{x_{i1}^T\bm{\beta_{1}}} + b_{i1} 
    = \beta_{H} + \beta^\omega_{H_i} %\mathbbm{1}_{H_i} 
    + \beta^\delta_{A_i} %\mathbbm{1}_{A_i} 
    + c_H + c'_{H} + b_{i1}, \\
    log(\mu_{i2}) %= \bm{x_{i2}^T\bm{\beta_{2}}} + b_{i2}
    = \beta_{A} + \beta^\omega_{A_i}%\mathbbm{1}_{A_i} 
    + \beta^\delta_{H_i}%\mathbbm{1}_{H_i} 
    + c_A + c'_{A} + b_{i2}.
\end{gather*}
\noindent Here, the random effects $\beta^\omega_{H_i}$ and $\beta^\delta_{H_i}$  capture the offensive and defensive strength, respectively, of the home team playing in game $i$. Similarly, $\beta^\omega_{A_i}$ and $\beta^\delta_{A_i}$ capture the offensive and defensive strengths of the away team playing at game $i$. The model also incorporates random-effects to identify a COVID-19 effect, $c_H$, and a post-COVID effect, $c'_H$. {\color{black} The intercepts, $\beta_H$ and $\beta_A$, represent the effects of playing at home and away, respectively. The HA effect is defined in the log scale as the rate of the average home team goals and the away goals $HA = log(\frac{\mu_1}{\mu_2}) = \beta_{H} -\beta_{A}$}.
In addition, we let the dispersion in the scores of both teams depend on the specific teams that are competing, and also on whether they play at home or away, by modeling the shape parameters as
\begin{gather*}
    log(\nu_{i1}) %= \bm{x_{i1}^T\bm\gamma_{i1}} 
    = \gamma_{H} + \gamma^\omega_{H_i}%\mathbbm{1}_{H_i} 
    + \gamma^\delta_{A_i}%\mathbbm{1}_{A_i} 
    + k_H + k'_{H}\\
    log(\nu_{i2}) %= \bm{x_{i1}^T}\bm{\gamma_{i2}} 
    = \gamma_{A} + \gamma^\omega_{A_i}%\mathbbm{1}_{A_i} 
    + \gamma^\delta_{H_i}%\mathbbm{1}_{H_i} 
    + k_A + k'_{A},
\end{gather*}
where the parameters $\gamma^\omega_{H_i}, \gamma^\delta_{A_i}, \gamma^\omega_{A_i},  \gamma^\delta_{H_i}, k_H,  k'_{H},  k_A, k'_{A}$ have similar interpretation as above.  We argue for this choice in the data applications below (see Figures \ref{fig:ID_soccer} and \ref{fig:ID_mlb}).

The vector of random effects, $\mathbf{b_i}=(b_{i1}, b_{i2}) $, is assumed to follow a multivariate normal centered at $0$ with an unrestricted covariance matrix $D$,
\begin{equation}
    \mathbf{b_i} | D \sim N_2(\bm 0, D), \quad i=1, \ldots, n,
\end{equation}
and models the within-game correlation between the two teams. More specifically, 
 we can prove that, under the model above, the covariance between the two team's observed counts, $y_{i1}$ and $y_{i2}$, is approximately 
\begin{eqnarray}
cov(y_{i1}, y_{i2}) \approx\, \lambda_{i1}\, e^{0.5 d_{11}}\, (e^{d_{12}}-1)\, \lambda_{i2}\, e^{0.5d_{22}},
\label{proof}
\end{eqnarray}
with $\lambda_{ij} = \exp(\bm{x_{ij}}^T\bm{\beta_j})$. The proof can be found in the Appendix. 
%\subsection{Prior distributions}
We complete the model by assuming standard independent priors on the vectors $(\bm\beta, \bm\gamma, D^{-1})$ as:
    \begin{align*}
    %\label{priors}
    \begin{split}
        \bm\beta &\sim N(\bm\beta_0, B_0^{-1}),\\
        \bm\gamma &\sim N(\bm\gamma_0, G_0^{-1}), \\
        D^{-1} &\sim Wishart(v_0, R_0),
        \end{split}
    \end{align*}
where $(\bm\beta_0, B_0, \bm\gamma_0, G_0, v_0, R_0)$ are fixed hyperparameters and $Wishart(v_0, R_0)$ denotes the Wishart distribution with $v_0$ degrees of freedom and scale matrix $R_0$. 

\subsection{Exchange algorithm and rejection sampler for the CMP}
\label{inference}
The posterior density of the proposed model is proportional to
    \begin{equation}
    \label{post}
        \Phi_2(\bm\beta|\bm\beta_0, B_0^{-1}) \Phi_2(\bm\gamma|\bm\gamma_0, G_0^{-1})f_W(D^{-1}|v_0, R_0) \prod_{i=1}^n p(\bm y_i|\bm\beta, \bm\gamma, b_i) \Phi_2(\bm b_i | \bm0, D), 
    \end{equation}
which is not available in closed form. Most importantly, in the CMP regression, Bayesian  inference for the parameters $\{\bm b_i\}$, $\bm\beta$, and $\bm\gamma$ is a \emph{doubly-intractable problem} as highlighted by \cite{murray2012}, and \cite{benson2021}. The likelihood is intractable, since it contains an intractable normalizing constant, $Z(\mu, \nu)$,  which is  parameter-dependent; thus, the posterior cannot be normalized either. Furthermore, since it is impossible to evaluate the likelihood point-wise, it is not possible to use standard MCMC methods for posterior inference. However, it is possible to use the Exchange algorithm devised by \cite{murray2012} to augment the doubly-intractable posterior with auxiliary latent variables, since it is possible to draw samples exactly from the intractable likelihood \citep{chanialidis2018, benson2021}. Indeed, 
\cite{chanialidis2018} showed that the CMP distribution is amenable to a rejection sampler. They proposed a first rejection sampler for the CMP without evaluating its normalizing constant, establishing an upper bound based on a piece-wise geometric distribution. 

In this paper, we implement a more efficient rejection algorithm proposed recently by \cite{benson2021} where they use two envelope distributions based on the value of the shape parameter $\nu$. More specifically, we construct a Markov chain procedure based on the Exchange algorithm to simulate from \eqref{post} using blocks of parameters $\{\bm b_i\}, \bm\beta, \bm\gamma$ and $D$ and full conditional distributions,
    \begin{equation}
    [\bm b|\bm y, \bm\beta, \bm\gamma, D];\; [\bm\beta | \bm y, \bm\gamma, \bm b];\; [\bm\gamma | \bm y, \bm\beta, \bm b];\; [D^{-1} | \bm b],       
    \end{equation}
where $\bm b = \{\bm b_1,...,\bm b_n\}$. The simulation output is obtained by recursively simulating these distributions using the updated values of the conditional variables at each step. More in detail, we obtain samples from these distributions as follows:

%To sample $b$ using the Metropolis-Hasting algorithm for example, we have that $p(b | y, \beta, \gamma, D) = \prod_{i = 1}^n{p(b_i | y_i, \beta, \gamma, D)}$, the product of n independent terms. Now, if we propose a move from each $b_i$ to $b_i^{*}$ using a proposal $h(b_i, b_i^*)$, then the acceptance ratio is
%    \begin{equation} \alpha_{MH}(b_i, b_i^*) = min \left \{ 1, \frac{\prod_{j = 1}^{J} p(y_{ij} \mid \beta_j, \gamma_j) h(b_i^*, b_i)\Phi_2(b_i^*|0,D)}{\prod_{j = 1}^{J} p(y_{ij} \mid \beta_j, \gamma_j) h(b_i^*, b_i)\Phi_2(b_i^*|0,D)} \right\} 
%    \end{equation}
    
%    \begin{equation} \alpha_{MH}(b_i, b_i^*) = min \left \{ 1, \frac{\prod_{j = i}^{J} \frac{q(y_{ij} | \beta_j, \gamma_j, b_{ij}^*)}{Z(\mu_{ij}^*,\nu_{ij})}h(b_i^*, b_i) \Phi_J(b_i^*|0,D)}{\prod_{j = 1}^J \frac{q(y_{ij} | \beta_j, \gamma_j, b_{ij})}{Z(\mu_{ij},\nu_{ij})}h(b_i,b_i^*) \Phi_J(b_i|0,D)} \right\} 
%    \end{equation}

%    $\alpha_{MH}$ involves the ratio $\frac{Z(\mu_{ij},\nu_{ij})}{Z(\mu_{ij}^*,\nu_{ij})}$ which makes its computation hard. Some possible solutions to this problem are the truncation of the normalizing constant, the use of asymptotic approximation \cite{shmueli2005}, and estimating upper and lower bounds for the value of the normalizing constant \cite{chanialidis2014}.

\begin{itemize}
\item {\bf Sampling $\bm b$:} The Exchange algorithm proposes to update the parameter vector $\bm b_i$ to $\bm b_i^*$, using a  Gaussian  proposal density centered at the current value, $h(\bm b_i, \bm b_i^*)$,  and augmenting the posterior with $n$ auxiliary draws,   $\bm y_i^* = (y_{i1}^*, y_{i2}^*)$ generated from the likelihood at the proposed parameter value, $\bm b_i^*$. 
The augmented posterior is
    \begin{equation}
        p(\bm b_i, \bm b_i^*, \bm y_i^*|\bm y_i) \propto p(\bm y_i |\bm b_i, \bm\beta, \bm\gamma) p(\bm b_i|D) h(\bm b_i,\bm b_i^*) p(\bm y_i^*|\bm b_i^*,\bm\beta,\bm\gamma).
    \end{equation}
    Therefore the acceptance ratio for the augmented posterior is
     \begin{equation} \alpha_{EA}(\bm b_i, \bm b_i^*) = \min \left \{ 1, \frac{\prod_{j = 1}^{2} \frac{q(y_{ij} | \bm\beta_j, \bm\gamma_j, b_{ij}^*)}{\cancel{Z(\mu_{ij}^*,\nu_{ij})}}h(\bm b_i^*, \bm b_i) \Phi_2(\bm b_i^*|\bm 0,D) \prod_{j = 1}^2 \frac{q(y_{ij}^* | \bm\beta_j, \bm\gamma_j, b_{ij})}{\cancel{Z(\mu_{ij},\nu_{ij})}}}{\prod_{j = 1}^{2} \frac{q(y_{ij} | \bm\beta_j, \bm\gamma_j, b_{ij})}{\cancel{Z(\mu_{ij},\nu_{ij})}}h(\bm b_i,\bm b_i^*) \Phi_2(\bm b_i|\bm 0,D) \prod_{j = 1}^2 \frac{q(y_{ij}^* | \bm\beta_j, \bm\gamma_j, b_{ij}^*)}{\cancel{Z(\mu_{ij}^*,\nu_{ij})}}} \right\}. 
    \end{equation}
    Since we can cancel out the normalizing constants, the acceptance ratio is
    \begin{equation}
    \alpha_{EA}(\bm b_i, \bm b_i^*) = \min \left \{ 1, \frac{\prod_{j = 1}^{2} q(y_{ij} | \bm\beta_j, \bm\gamma_j, b_{ij}^*)\Phi_2(\bm b_i^*|\bm 0,D) \prod_{j = 1}^2 q(y_{ij}^* | \bm\beta_j, \bm\gamma_j, b_{ij})}{\prod_{j = 1}^2 q(y_{ij} | \bm\beta_j, \bm\gamma_j, b_{ij}) \Phi_2(\bm b_i|\bm 0,D) \prod_{j = 1}^2 q(y_{ij}^* | \bm\beta_j, \bm\gamma_j, b_{ij}^*)} \right\}.   
    \end{equation}

\item {\bf Sampling $\bm \beta$ and $\bm \gamma$:} we need to sample $\bm\beta$ given $(\bm b, \bm\gamma, D)$ and $\bm\gamma$ given $(\bm b, \bm\beta, D)$. We can see that the posterior of $\bm\beta$ and $\bm\gamma$ is proportional to 
    \begin{equation*}
    \begin{split}
        p(\bm\beta,\bm\gamma|y,b,D) \propto \phi_k(\bm\beta|\bm\beta_0, B_0^{-1}) \phi_k(\bm\gamma|\bm\gamma_0, G_0^{-1}) \prod_{j=1}^2{p(\bm y_{\cdot j}|b_{\cdot j}, \bm\beta_j, \bm\gamma_j)},
        \end{split}
    \end{equation*}
where $p(\bm y_{\cdot j}|\bm\beta_j,\bm\gamma_j,b_{ij}) = \prod_{i=1}^n p(y_{ij}|\bm\beta_j, \bm\gamma_j, b_{ij})$.
We sample each of the $\bm\beta_j$'s and $\bm\gamma_j$'s by implementing a sequence of Exchange algorithms for each component of the parameter vectors. % Each component of $\bm\beta$, $\bm\beta_j$'s, and each component of $\bm\gamma$, $\bm\gamma_j's$, are revised one at a time.
 More specifically, the proposal $\bm\beta_j^*$ is given as proposed in the Robust adaptive Metropolis algorithm (RAM) \cite{vihola2012robust}, using a Gaussian proposal density. Then we draw $n$ auxiliary data ${\bm y_{\cdot j}}^*$, and the acceptance rates for $\bm \beta$ are
\begin{equation}
    \alpha_{{EA}_j}(\bm\beta_j, \bm\beta_j^*) = \min \left \{ 1, \frac{\phi_{k_j}(\bm\beta_j^*|\bm\beta_{0j}, B_{0j}^{-1}) q(\bm y_{\cdot j} | \bm\beta_j^*, \bm\gamma_j, \bm b_{\cdot j}) q(\bm y_{\cdot j}^*|\bm\beta_j, \bm\gamma_j, \bm b_{\cdot j})}{\phi_{k_j}(\bm\beta_j|\bm\beta_{0j}, B_{0j}^{-1}) q(\bm y_{\cdot j} | \bm\beta_j, \bm\gamma_j, \bm b_{\cdot j}) q(\bm y_{\cdot j}^*|\bm\beta_j^*, \bm\gamma_j, \bm b_{\cdot j})} \right\}, j \leq 2.
\end{equation}
Similarly, we have the sequence of acceptance ratios for $\gamma$ as
\begin{equation}
    \alpha_{{EA}_j}(\bm\gamma_j, \bm\gamma_j^*) = \min \left \{ 1, \frac{\phi_{k_j}(\bm\gamma_j^*|\bm\gamma_{0j}, G_{0j}^{-1}) q(\bm y_{\cdot j} | \bm\beta_j, \bm\gamma_j^*, \bm b_{\cdot j}) q(\bm y_{\cdot j}^*|\bm\beta_j, \bm\gamma_j, \bm b_{\cdot j})}{\phi_{k_j}(\bm\gamma_j|\bm\gamma_{0j}, G_{0j}^{-1}) q(\bm y_{\cdot j} | \bm\beta_j, \bm\gamma_j, \bm b_{\cdot j}) q(\bm y_{\cdot j}^*|\bm\beta_j, \bm\gamma_j^*, \bm b_{\cdot j})} \right\}, j \leq 2.
\end{equation}

\item {\bf Sampling $D$:} the full conditional for $D$ is a Wishart distribution $p(D|\bm b) \propto p(\bm b|D)p(D) = \prod_{i=i}^n \phi_2(\bm b_i|\bm 0,D)f_w(D^{-1}|v_0, R_0^{-1})$. Thus $D^{-1}|\bm b \sim Wishart\left(n+v_0, \left[R_0^{-1} + \sum_{i=1}^n(\bm b_i\bm b_i^T)\right]^{-1}\right)$, and is sampled directly.
%% Summary?    

\end{itemize}
\section{Simulation Study}
\label{simulations}

{\color{black}
\subsection{Data generation and parameter settings}

This section aims at illustrating and evaluating the performance of the proposed CMP model under simulated data on different scenarios. More specifically, we assess how the model fits score data and compare performances with the widely employed independent Poisson and Negative Binomial models. Additionally, we test the capability of the proposed model to accurately capture the true HA effect in scenarios with general over-, equi-, and under-dispersed data.

Based on the model presented in Section \ref{model}, we fix the home advantage (HA) effect by setting $\beta_H = 0.6$ and $\beta_A = 0.1$, resulting in a true effect of $HA = 0.5$. We simulate scores within a framework comprising 20 teams, labeled as Team 1 through Team 20, across various seasons. Each season consists of a total of 380 games, where every team competes against each other, both at home and away. In this simulation setup, we did not consider the effect of the COVID-19 season.
We assign to every team specific strength attacking $(\beta^\omega)$ and defending $(\beta^\delta)$, distinguishing between their performance at home (H) and away (A). At home, attacking strengths range from 0.5 for Team 1 to -0.5 for Team 20. The attacking strengths playing away range from -0.42 to 0.42, with a notable exception for Teams 6 through 15, where the sign is inverted. The list of the effects for every team can be found in Table \ref{tab:sim_strengths}. This approach categorizes the teams into four groups: the first group (Teams 1:5) is characterized by strong performance both at home and away. The second group (Teams 6:10) demonstrates above-average attacking strength at home but below-average performance away. Conversely, the third group (Teams 11:15) displays below-average strength when playing at home but stronger away attacking capabilities. Finally, the last group (Teams 16:20) possesses the lowest strengths compared to the other teams and is, therefore, composed of the worst performing teams. As for the strengths defending, they were assigned from 0.5 to -0.5 at Home and from 0.6 to -0.6 playing Away.

\begin{table}[ht]
\caption{Simulation study: Team's strengths at Attack and Defense playing Home or Away}
    \centering
    \resizebox{\columnwidth}{!}{%
    \begin{tabular}{l|rrrrr|rrrrr|rrrrr|rrrrr|}
      & \multicolumn{5}{c}{Good attack at home and away} & \multicolumn{5}{c}{Good at home, bad away} & \multicolumn{5}{c}{Bad at home, good away} & \multicolumn{5}{c}{Bad at home, bad away} \\
     Team & 1 & 2 & 3 & 4 & 5 & 6 & 7 & 8 & 9 & 10 & 11 & 12 & 13 & 14 & 15 & 16 & 17 & 18 & 19&20 \\ 
      \hline
    $\beta_H^\omega$ & 0.50 & 0.45 & 0.40 & 0.34 & 0.29 & 0.24 & 0.18 & 0.13 & 0.08 & 0.03 & -0.03 & -0.08 & -0.13 & -0.18 & -0.24 & -0.29 & -0.34 & -0.40 & -0.45 & -0.5\\ 
      $\beta_H^\delta$ & 0.50 & 0.45 & 0.40 & 0.34 & 0.29 & 0.24 & 0.18 & 0.13 & 0.08 & 0.03 & -0.03 & -0.08 & -0.13 & -0.18 & -0.24 & -0.29 & -0.34 & -0.40 & -0.45 & -0.5\\ 
      $\beta_A^\omega$ & 0.42 & 0.38 & 0.34 & 0.29 & 0.25 & -0.20 & -0.16 & -0.11 & -0.07 & -0.02 & 0.02 & 0.07 & 0.11 & 0.16 & 0.20 & -0.25 & -0.29 & -0.34 & -0.38 & -0.42\\ 
      $\beta_A^\delta$ & 0.60 & 0.54 & 0.47 & 0.41 & 0.35 & 0.28 & 0.22 & 0.16 & 0.10 & 0.03 & -0.03 & -0.10 & -0.16 & -0.22 & -0.28 & -0.35 & -0.41 & -0.47 & -0.54 & -0.60\\ 
    \end{tabular}
    }
    \label{tab:sim_strengths}
    \end{table}
    
As for the shape parameters $\gamma$, we consider that strongest teams have a larger dispersion in the number of goals, so their shape parameters will be larger than weakest teams, then we set them sequentially from 0.35 for Team 1 to -0.35 for Team 20, both at home and away. We consider three scenarios where data tend to be over-, equi- or under-dispersed based on the value of the intercept parameters $\gamma_H$ and $\gamma_A$. In the first scenario, we let $\gamma_H = \gamma_A = 0$, in which we observe that the shape parameters $\nu_{.1}$, $\nu_{.2}$ are centered around 1, corresponding to equi-dispersed data. In the second scenario, we set $\gamma_H = \gamma_A = 0.5$, resulting in shape parameters mainly greater than 1, indicating an under-dispersed scenario. In the last scenario, we consider $\gamma_H = \gamma_A = -0.5$, which corresponds to data mostly over-dispersed. 

We fit the CMP model assuming non-informative priors by setting the hyper-parameters to $\beta_0 = \gamma_0 = 0$, $B_0 = G_0 = 0.1I$, $\nu_0 = 50$, $R_0 = I$. A motivation for this choice of values and a prior sensitivity analysis can be found in the Appendix. We run MCMC chains with 30K samples in total, where 10K are used for burn-in. We tune the MCMC algorithm to achieve an acceptance ratio of approximately 40\%. For each scenario, we simulate $40$ replicated datasets, fit the three models, estimate the HA effect, and calculate the Deviance Information Criterion (DIC) \citep{spiegelhalterbayesian} to assess the goodness of fit, as also done by \citet{chanialidis2018}. In the case of the Conway-Maxwell-Poisson model, direct computation of the DIC is not possible due to the intractable constant in the likelihood function. Thus, we adopt an approach suggested by \citet{benson2021} to utilize an unbiased likelihood estimator that relies on the number of acceptances ($r$) obtained from the rejection sampling algorithm used in the Exchange algorithm,

\begin{equation*}
	\widehat{f}^{(r)}(y|\theta) = \prod_{i = 1}^n{\frac{q_f(y_i | \theta_i)}{Z_g(\phi_i)}\frac{{\widehat{M}^{(r)}}_{f/g,i}}{B_{f/g,i}}},
\end{equation*}

\noindent where ${Z_g}$ is the normalizing constant of the envelope distribution, $B_{f/g}$ is a tractable enveloping bound, and $\widehat{M}^{(r)}$ is the number of draws required for $r$ acceptances.  For more detailed information, please refer to \cite{benson2021}. They recommend a high value for the number of acceptances ($r$). In our case, we selected $r = 1000$. For computational simplicity we calculate the DIC using 100 random samples from the MCMC algorithm.

\subsection{Results}
Figure \ref{fig:simulation_ha} shows the estimation and the $95\%$ credible interval of the HA effect under the models considered across 1, 3 and 5 seasons. Looking at those results, we observe that, by modeling the dispersion parameter, we indeed capture the true effect on almost all the scenarios, with improved precision for cases with larger sample sizes. In contrast, under the Negative Binomial and the Poisson models, the effect is not captured if the scores tend to be over or under-dispersed. These results highlight the fact that our model is robust in fitting data exhibiting any type of dispersion.

\begin{figure}[!ht]
   \centering
   \includegraphics[scale = 0.5]{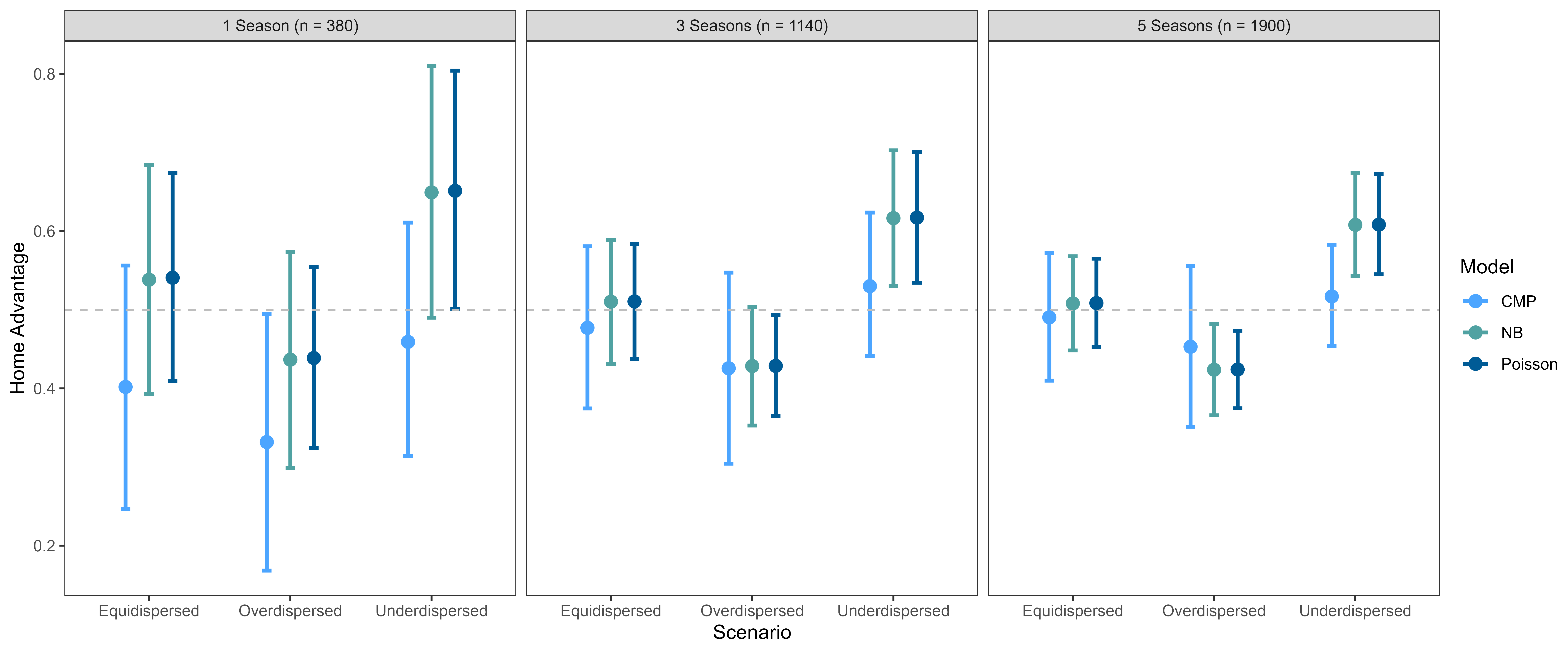}
   \caption{Simulation study: comparison of HA effect estimation and their 95\% credible intervals. Dashed line corresponds to the true value}
   \label{fig:simulation_ha}
\end{figure}

Deviance Information Criterion (DIC) values for the different models averaged across replicated datasets are shown in Table \ref{tab:simulation_ha}. Both the Conway-Maxwell-Poisson and Poisson models demonstrate reasonable performance in capturing equi-dispersion, with no significant difference observed between the DIC values for the two response variables in cases with small sample size. In the over-dispersed scenarios, the Negative Binomial model generally outperforms the Poisson model but shows comparable or worse performance compared to the Conway-Maxwell-Poisson model. As the sample size increases to $n = 1900$, the Conway-Maxwell-Poisson model exhibits the lowest DIC values, and significant differences among the models are observed in both over- and under-dispersed scenarios. These results suggest that the Conway-Maxwell-Poisson model outperforms the Poisson and Negative Binomial models when dealing with large sample sizes. Furthermore, the Conway-Maxwell-Poisson model adapts very well to various dispersion scenarios.

\begin{table}[ht]
\caption{Simulation study: DIC (Mean, SD) across 40 replicates. Bolded numbers represent the lowest values for each response in the bivariate outcome model.}
\resizebox{\textwidth}{!}{%
\begin{tabular}{@{}cl|ll|ll|ll@{}}
 & \multicolumn{1}{l}{} & \multicolumn{2}{c}{1 Season $(n = 380)$} & \multicolumn{2}{c}{3 Season $(n = 1140)$} & \multicolumn{2}{c}{5 Seasons $(n = 1900)$} \\
Data & \multicolumn{1}{c}{Model} & \multicolumn{1}{c}{$y_1$} & \multicolumn{1}{c}{$y_2$} & \multicolumn{1}{c}{$y_1$} & \multicolumn{1}{c}{$y_2$} & \multicolumn{1}{c}{$y_1$} & \multicolumn{1}{c}{$y_2$} \\
\midrule
\multirow{3}{*}{Equi} & CMP & 1355.91 (35.02) & \textbf{1091.58 (31.28)} & \textbf{3922.93 (60.83)} & \textbf{3220.48 (55.66)} & \textbf{6487.45 (80.34)} & \textbf{5343.95 (73.92)} \\
 & Poisson & \textbf{1333.79 (31.86)} & 1104.65 (32.39) & 3932.85 (60.14) & 3245.29 (52.67) & 6542.70 (63.86) & 5390.63 (69.82) \\
 & NB & 1349.06 (28.36) & 1118.44 (30.92) & 3937 (53.56) & 3252.55 (50.09) & 6538.25 (58.19) & 5392.93 (67.41) \\
 \midrule
\multirow{3}{*}{Over} & CMP & \textbf{1490.37 (30.92)} & \textbf{1251.37 (28.85)} & \textbf{4391.69 (54.74)} & \textbf{3724.38 (50.12)} & \textbf{7269.45 (73.87)} & \textbf{6194.46 (73.13)} \\
 & Poisson & 1509.55 (37.25) & 1279.62 (34.64) & 4493.19 (75.45) & 3805.90 (56.38) & 7470.95 (99.78) & 6340.21 (81.99) \\
 & NB & 1498.13 (30.25) & 1280.05 (31.38) & 4407.10 (58.35) & 3754.65 (45.27) & 7314.27 (77.73) & 6238.71 (64.75) \\
 \midrule
\multirow{3}{*}{Under} & CMP & 1236.08 (24.67) & \textbf{962.93 (26.79)} & 3523.70 (46.89) & \textbf{2810.50 (55.05)} & \textbf{5722.46 (62.80)} & \textbf{4596.15 (72.28)} \\
 & Poisson & \textbf{1203.57 (18.45)} & 966.65 (28.49) & \textbf{3522.18 (38.23)} & 2817.38 (40.78) & 5841.81 (39.09) & 4665.59 (54.50) \\
 & NB & 1228.44 (17.73) & 985.75 (28.58) & 3555.02 (36.92) & 2840.20 (40.07) & 5880.47 (37.91) & 4692.64 (53.74) \\ 
\end{tabular}%
}
\label{tab:simulation_ha}
\end{table}
}

\label{data_analysis}
\section{Analysis of Soccer English Premier League Data}
\label{data_analysis_soccer}
\subsection{Data description}
We illustrate the performance of our model on sports count data, by using game outcomes from the English Premier League over the span of three seasons: 2019-2020, 2020-2021, and a portion of 2021-2022.\footnote{Data extracted from: \url{https://www.football-data.co.uk/}} The dataset comprises a total of 1237 games. On average, we observe 1.49 goals per game for the home team, with a corresponding variance of 1.79. Similarly, the away team records an average of 1.27 goals per game, accompanied by a variance of 1.49. These statistics indicate that both the home and away goals exhibit a slightly higher variance than their respective means. To assess the relationship between the mean and the variance, we utilize the dispersion statistic, denoted as $\sigma_p$. This measure takes into account the sample size and model complexity while quantifying the difference between the variance and the mean. Notably, for both the home and away goals, $\sigma_p$ is found to be less than $1.04$, indicating no significant deviation between the mean and the variance. It is worth noting that a threshold of $\sigma_p > 1.2$ is typically employed to identify over-dispersion \citep{payne2018empirical}. Hence, the results suggest that the Poisson model may be a reliable choice for modeling this data.

\begin{figure}[!ht]
    \caption{Analysis of Premier League data: Distribution of the scores of the 1237 games in the English Premier League}
    \centering
    \includegraphics[scale = 0.5]{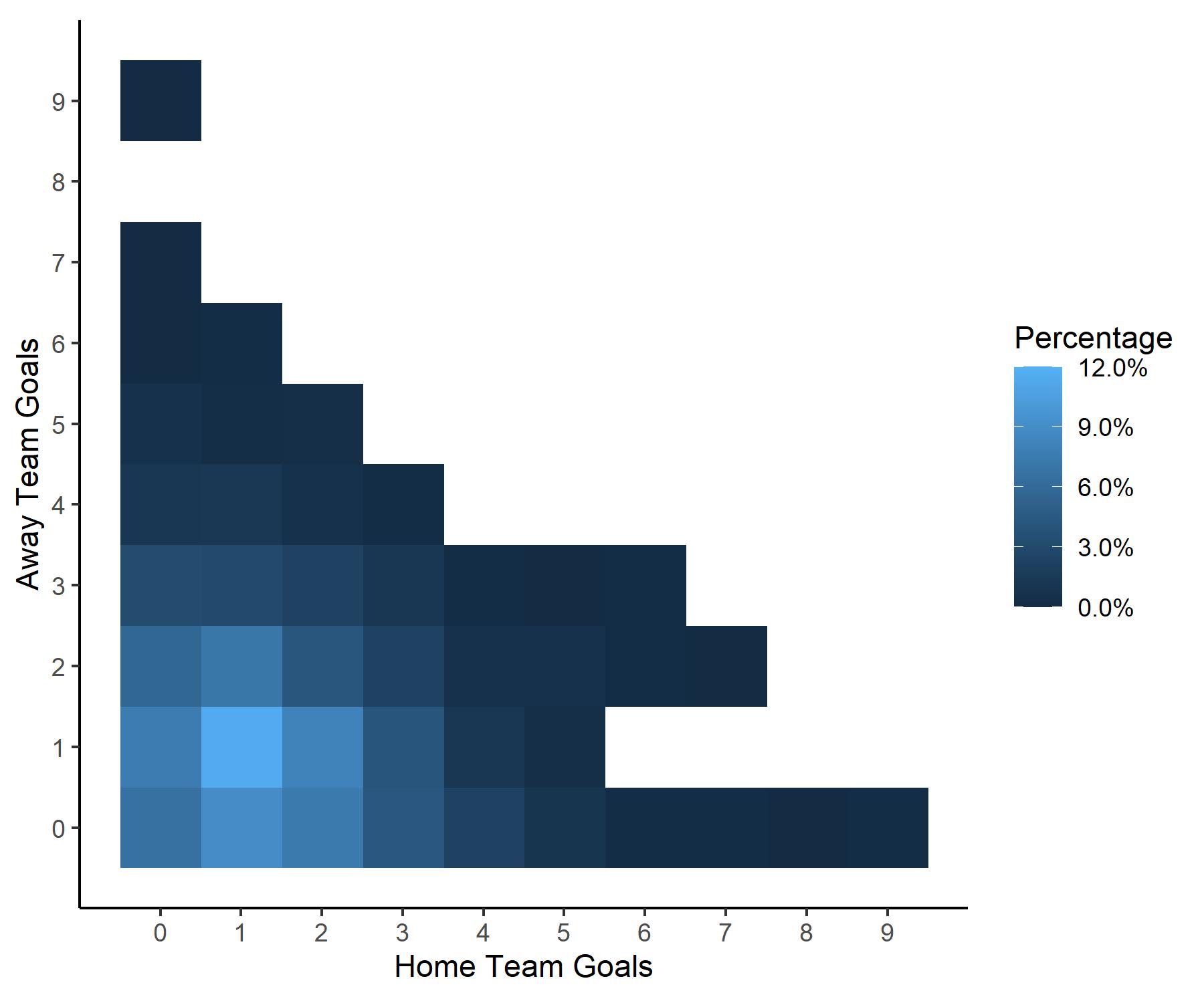}
    \label{fig:dist_scores_soccer}
\end{figure}

The negative Spearman's correlation coefficient of -0.143 suggests a slight monotonic relationship between the goals scored by the home and away teams. This finding is further supported by the distribution of goals depicted in Figure \ref{fig:dist_scores_soccer}. Specifically, the relative frequency of the home team scoring more goals than the away team ($Fr($Home Team Goals $ > $ Away Team Goals$)$) is found to be 0.429, while the relative frequency of the home team scoring fewer goals than the away team ($Fr($Home Team Goals $ < $ Away Team Goals$)$) is 0.34. These findings underscore the presence of a correlation effect between the goals scored by the home and away teams. Therefore, it is important to consider this relationship when analyzing the data.

\begin{figure}[!ht]
   \centering
    \includegraphics[scale = 0.06]{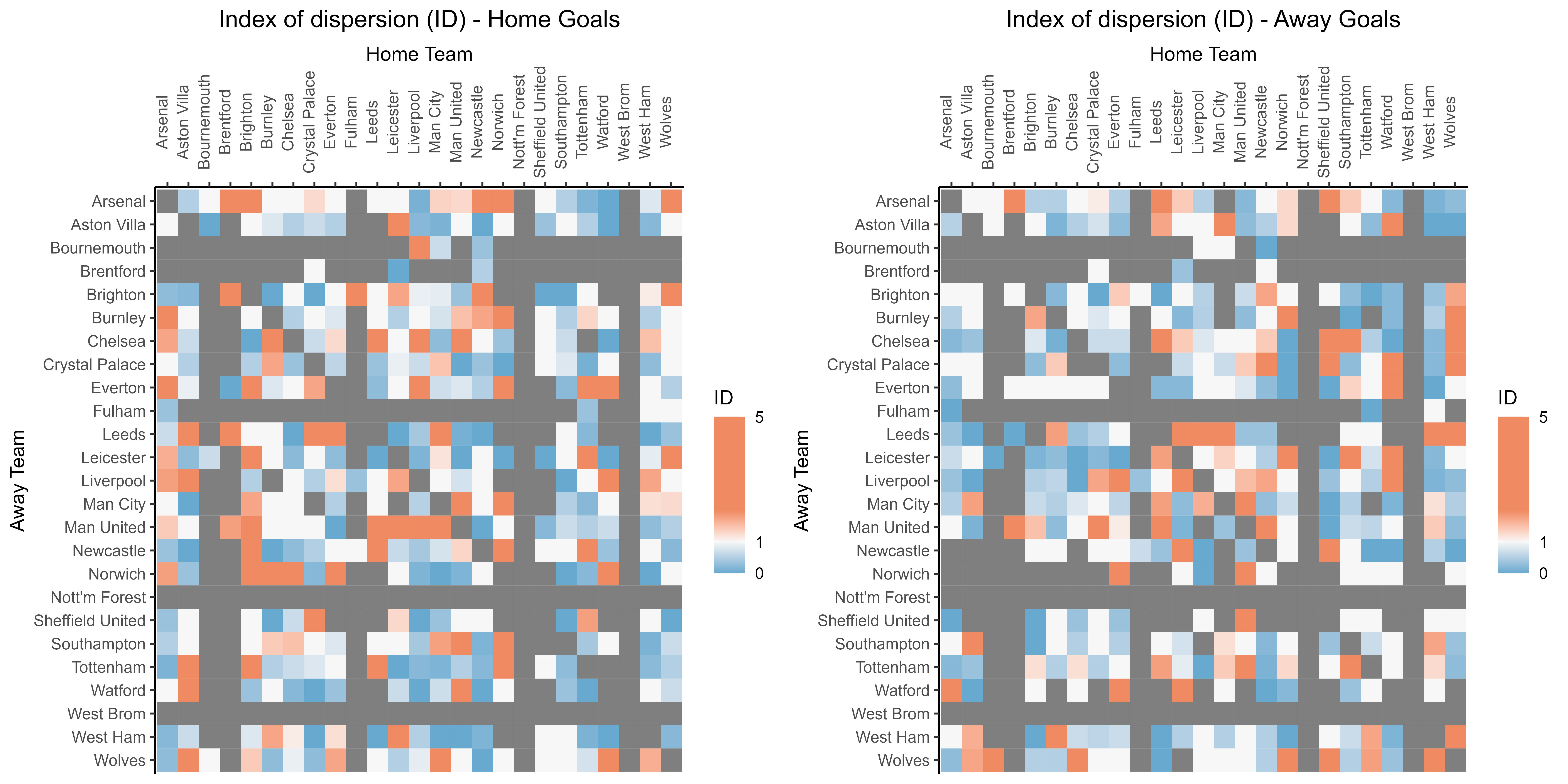}
    \caption{Analysis of Premier League data: Indices of dispersion (ID)}
    \label{fig:ID_soccer}
\end{figure}

To complement the analysis of dispersion, in Figure \ref{fig:ID_soccer} we display the Index of dispersion (ID) for the Premier League data. The ID or Variance to Mean Ratio (VMR) is defined as the ratio between the sample variance and the sample mean, $\frac{s^2}{\bar{x}}$. If data tend to be equi-dispersed this value is around to 1, while for over-dispersed data will be greater than 1. 
The heat map on the left of Figure \ref{fig:ID_soccer} corresponds to the ID of the goals scored by the Home Team, while the heat map on the right is the ID of the goals scored by the Away team. We can observe that the index of dispersion is generally close to 1 for specific games with over-dispersion and others with under-dispersion, depending on the teams playing. Additionally, the two heatmaps are different, suggesting that the effect on the dispersion of the home goals and the away goals made by the teams is different when they play at home than when they play on the road. For instance, when Arsenal plays at home, the ID of the goals they scored (first column, left heatmap) tends to be greater than 1, but the goals they conceded (first column, right heatmap) tend to be more under-dispersed. Similarly we can interpret the values for Everton. When Everton plays away, the ID of the number of goals scored by the opponent (Everton row, left heatmap) varies depending on the opponent; 7 out of them are over-dispersed. However, the ID of the goals they score tends to be more equi-dispersed (Everton row, right heatmap). 

\begin{figure}[!ht]
    \centering
    \subfigure[Average goals per game playing at home field]{\includegraphics[width=0.45\textwidth]{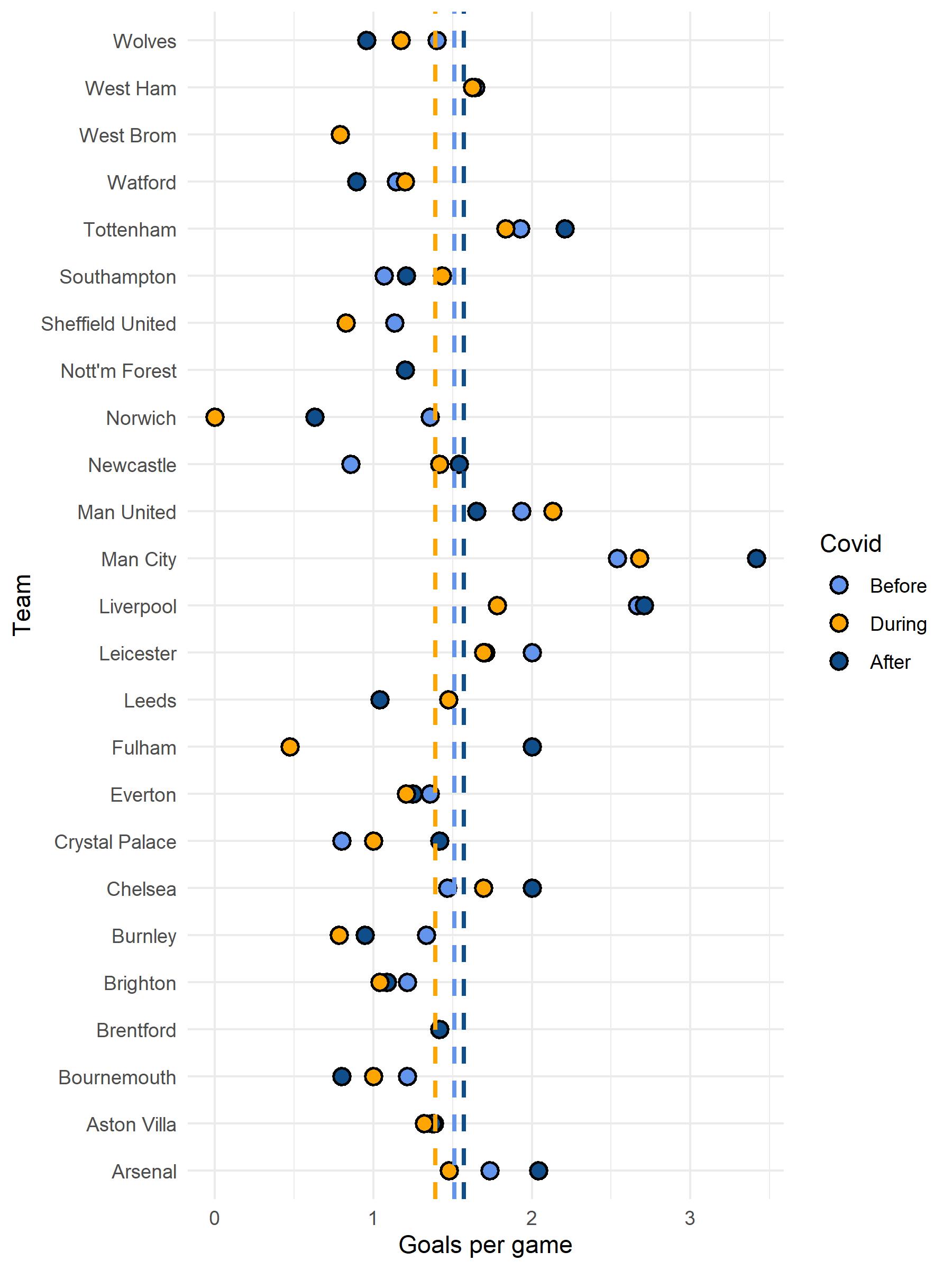}
    \label{fig:avg_home}}
    \hspace{0.05\textwidth} % Add horizontal space between the subfigures
    \subfigure[Average goals per game playing at away field]{\includegraphics[width=0.45\textwidth]{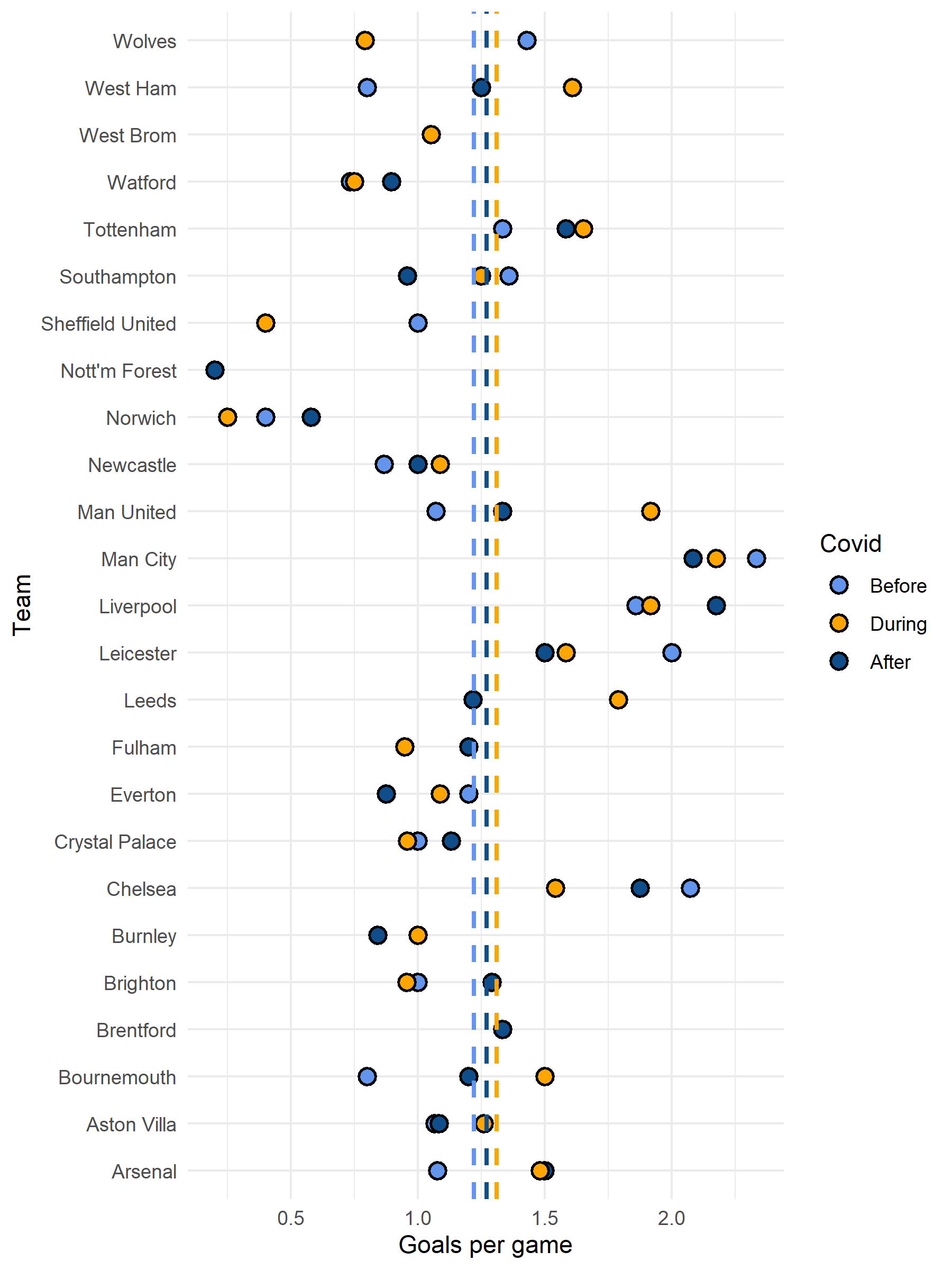}
    \label{fig:avg_away}}
    \caption{Analysis of Premier League data: Average team goals per game by Covid season}
    \label{fig:avg_goals}
\end{figure}

Figure \ref{fig:avg_goals} shows the average team's goals per game during the COVID-19 restrictions. We observe a decrease in the number of goals scored per game. Prior to the pandemic, the average rate of goals scored by home teams per game was 1.51. However, this figure decreased to 1.38 during the pandemic and subsequently increased to 1.57 after the restrictions were lifted. A detailed breakdown by teams can be found in Figure \ref{fig:avg_goals} (a). From this plot, we can discern that nearly half of the teams experienced a reduction in the average number of goals scored per game during the pandemic, followed by an increase afterwards.
Conversely, the average number of goals scored by away teams per game exhibited a small increase during the COVID-19 pandemic, rising from 1.22 before the pandemic to 1.31 during the restrictions. However, this trend reversed after the COVID-19 restrictions were lifted, resulting in a decrease to 1.27 goals per game. Figure \ref{fig:avg_away} (b) provides a visual representation of these observations.

Furthermore, in Figure \ref{fig:dist_goals_covid} we can examine the difference between the number of goals scored by the home team and the away team and its distribution across the COVID-19 seasons under consideration. A negative difference indicates that the away team scored more goals than the home team, signifying an away team victory, while a positive difference implies that the home team scored more goals, representing a home team victory. We see that the distribution of the home goals difference remains similar before and after the COVID-19 restrictions. However, during the pandemic restrictions, there is a slight increase in the proportion of games with a negative difference. From these findings, it is apparent that the percentage of home team wins significantly decreased during the COVID-19 restrictions (\% Home Wins: Before - $45\%$, After - $44\%$, During - $39\%$), while the percentage of home team losses increased  (\% Home Losses: Before - $31\%$, After - $31\%$, During - $39\%$). This observation supports the hypothesis that the HA diminished during the pandemic and was recovered afterwards.
However, an ANOVA test to ascertain whether there were significant differences in the home goals difference across the stages of COVID-19 (before, during, and after the restrictions) yielded a p-value of 0.154, indicating no significant evidence to reject the null hypothesis that the average home goals difference per game did not change during the various stages of COVID-19 considered. 

\begin{figure}[!ht]
    \centering
    \includegraphics[scale = 0.6]{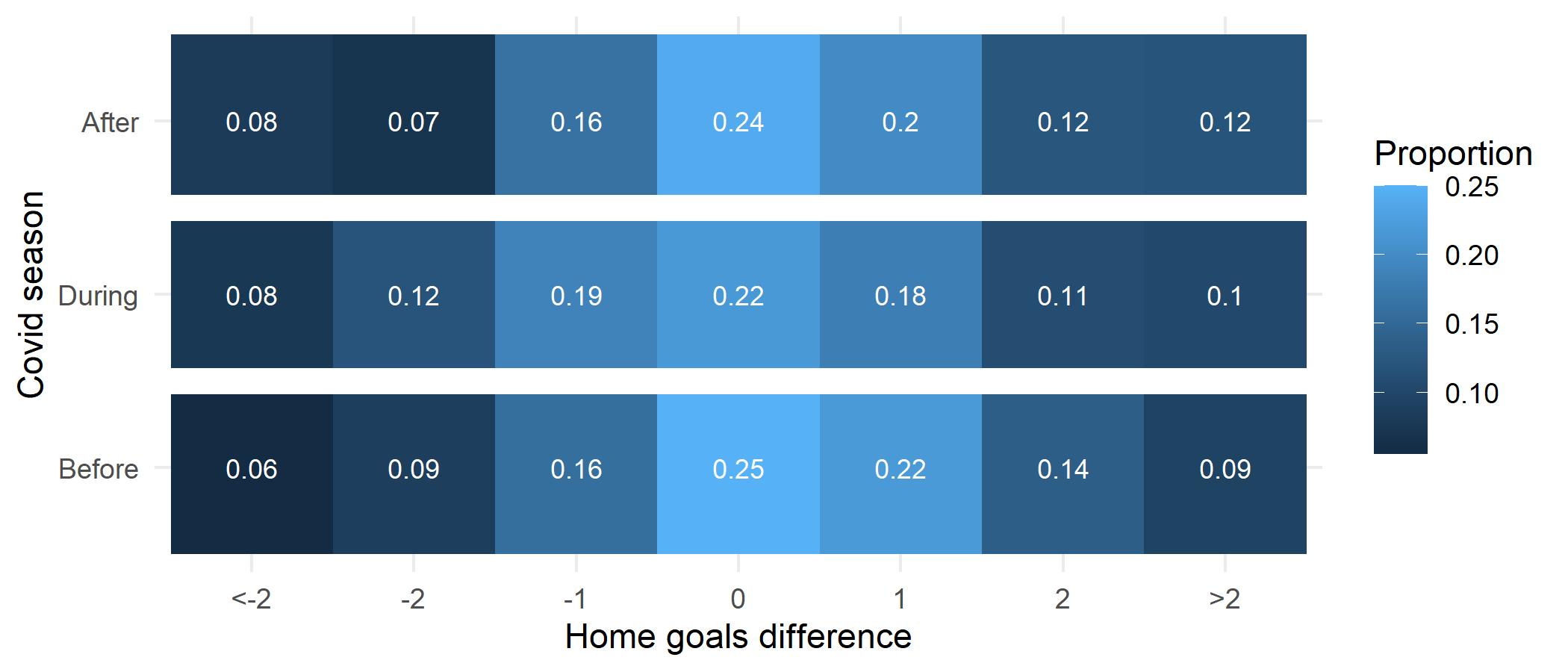}
    \caption{Analysis of Premier League data: distribution of the difference between Home goals and Away goals}
    \label{fig:dist_goals_covid}
\end{figure}

\subsection{Model fitting}

We fit the proposed CMP model on the Premier League data, and then  compare the performance with those of the Negative Binomial and Poisson models.  Given the negative correlation between the two responses and the apparent equi-dispersion of the data, our model appears to be a suitable approach for modeling the Premier League data and identifying any variations in the HA during the crowd restrictions imposed amid the COVID-19 pandemic.

\begin{figure}[!ht]
    \centering
    \includegraphics[scale = 0.6]{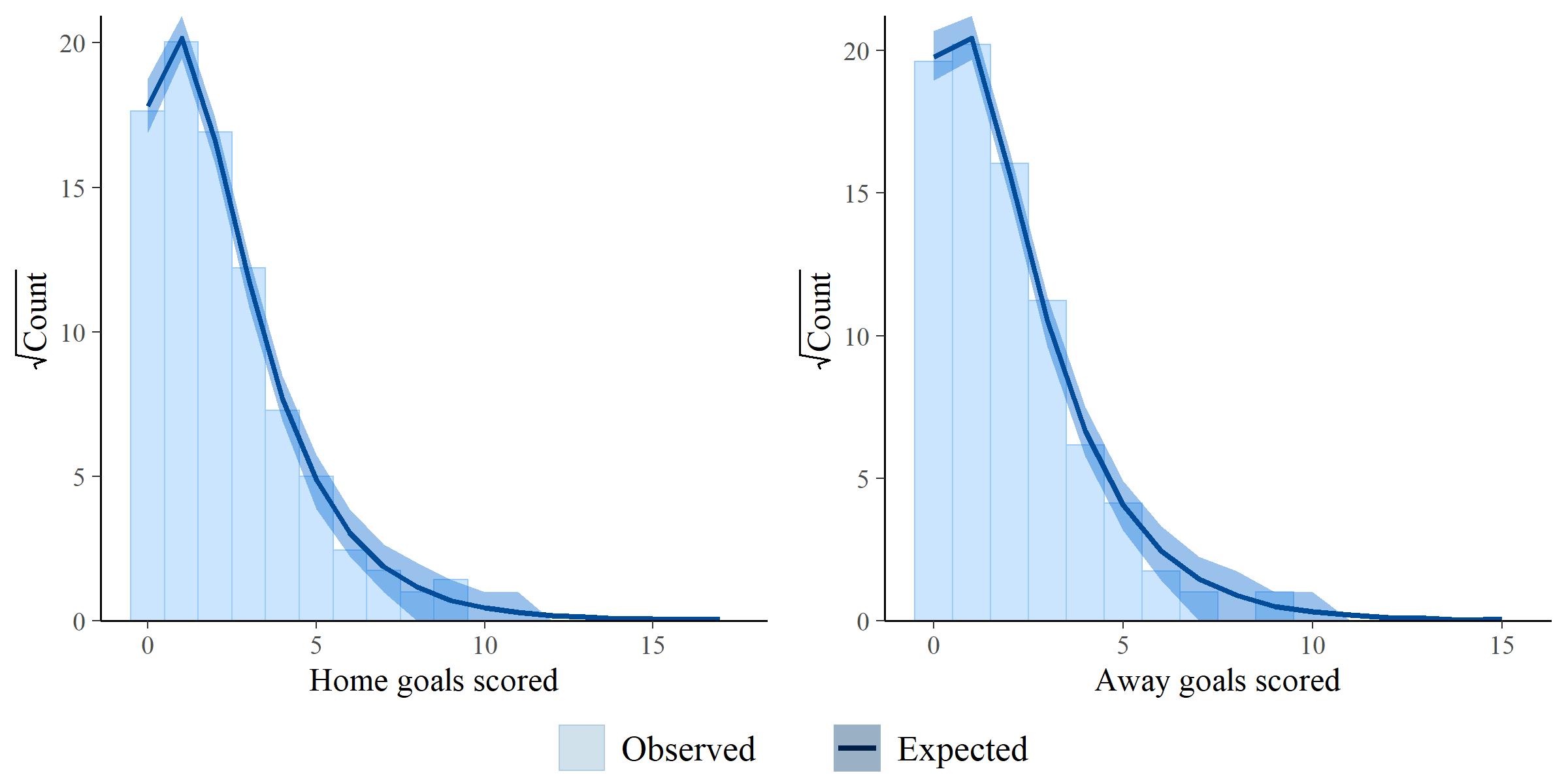}
    \caption{Analysis of Premier League data: empirical distribution and the posterior predictive distribution of the home goals and away goals}
    \label{fig:fit_soccer}
\end{figure}

For our analysis, we assume the following prior hyperparameters: $\beta_0 = \gamma_0 = 0$, $B_0 = G_0 = 0.1\, I$, $\nu_0 = 50$, and $R_0 = I$,  to reflect minimal prior knowledge. {\color{black} In this case, we used 180K MCMC iterations with a burn-in of 50K iterations, which achieved an effective sample size greater than 500 for the estimated parameters}. Figure \ref{fig:fit_soccer} displays a comparison between the empirical distribution of the data and the posterior predictive distribution obtained from our model. It is evident that our model effectively captures the distribution of the data, suggesting a good fit between the observed and the predicted values.

\begin{figure}[!ht]
    \centering   
    \includegraphics[scale = 0.4]{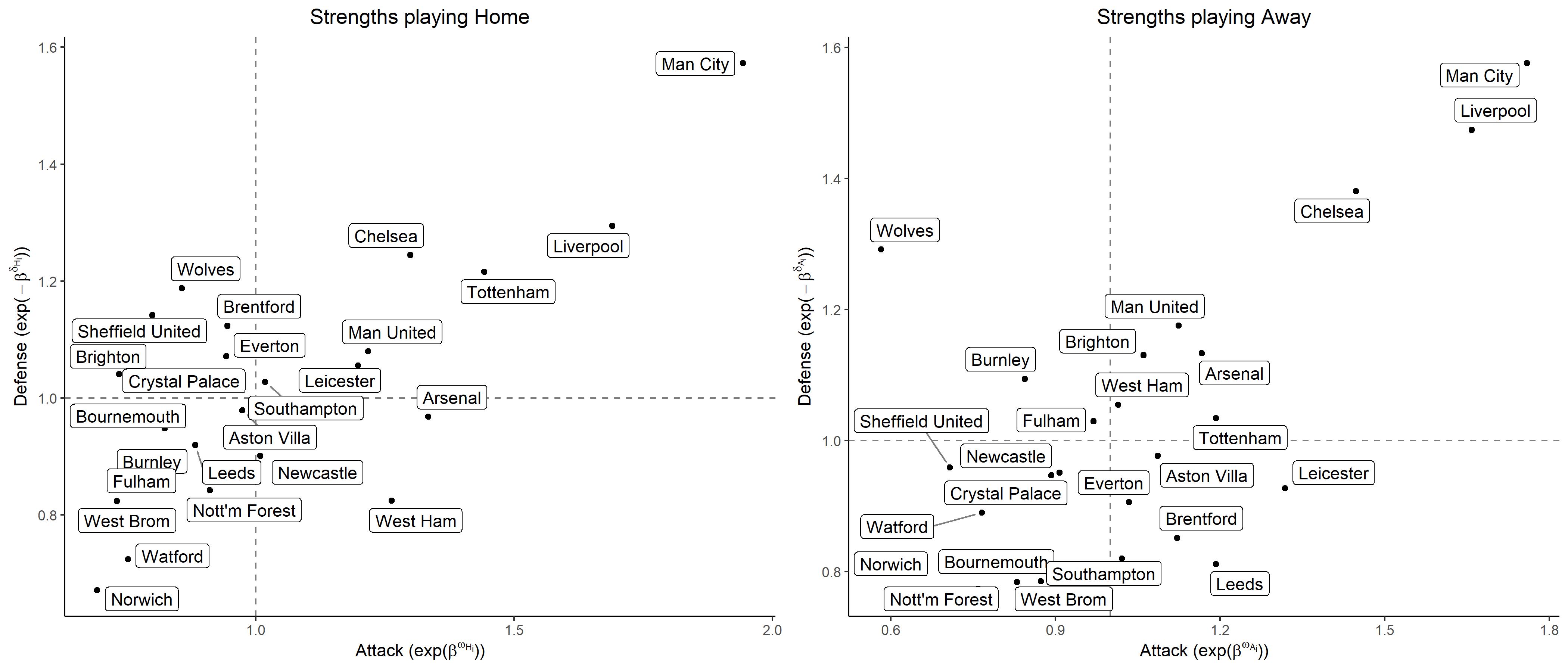}
        \caption{Analysis of Premier League data: Attack and defense strength of each team}
    \label{soccer_posteriors}
\end{figure}

Figure \ref{soccer_posteriors} presents a comparison of the attack and defense strengths of each team based on the posterior estimations. {\color{black} Higher values in the defense parameters of any team $\beta^\delta$ indicate that it is more likely to concede goals. We plot the exponential of the negative parameter such that teams positioned above the vertical and horizontal lines at 1 exhibit better performance than the team average}. Notably, Manchester City and Liverpool, the champions in recent seasons, demonstrate the strongest attack and defense, whether playing at home or away. Interestingly, there are teams that seem to benefit from playing away, as indicated by their superior attack strength compared to other teams. West Ham, Brighton, and Leeds are notable examples. {\color{black} This is an intriguing finding because these teams were actually battling to avoid relegation during the seasons considered in the study. In an interview in \textit{The Athletic}, Shaka Hislop, former West Ham goalkeeper, suggested a possible reason for this phenomenon. He points out that teams facing difficulties often find it easier to play away from home due to reduced pressure. Hislop notes that the intense scrutiny and expectations of home crowds can inhibit players' performance, whereas playing away allows for more freedom and different strategies, that could contribute to improved results \citep{Thomas2019-jd}.} On the other hand, several teams demonstrate a lower number of goals conceded when playing at home compared to playing away. Noteworthy teams in this regard include Everton, Crystal Palace, and Brighton. These findings provide valuable insights into the varying performance and strategies of different teams in terms of attack and defense strengths, shedding light on the dynamics of the Premier League. {\color{black}Similarly, we can interpret the estimated shape parameters. A detailed description and this estimation can be found in the Appendix.}

\begin{figure}[!ht]
    \centering
    \includegraphics[scale=0.45]{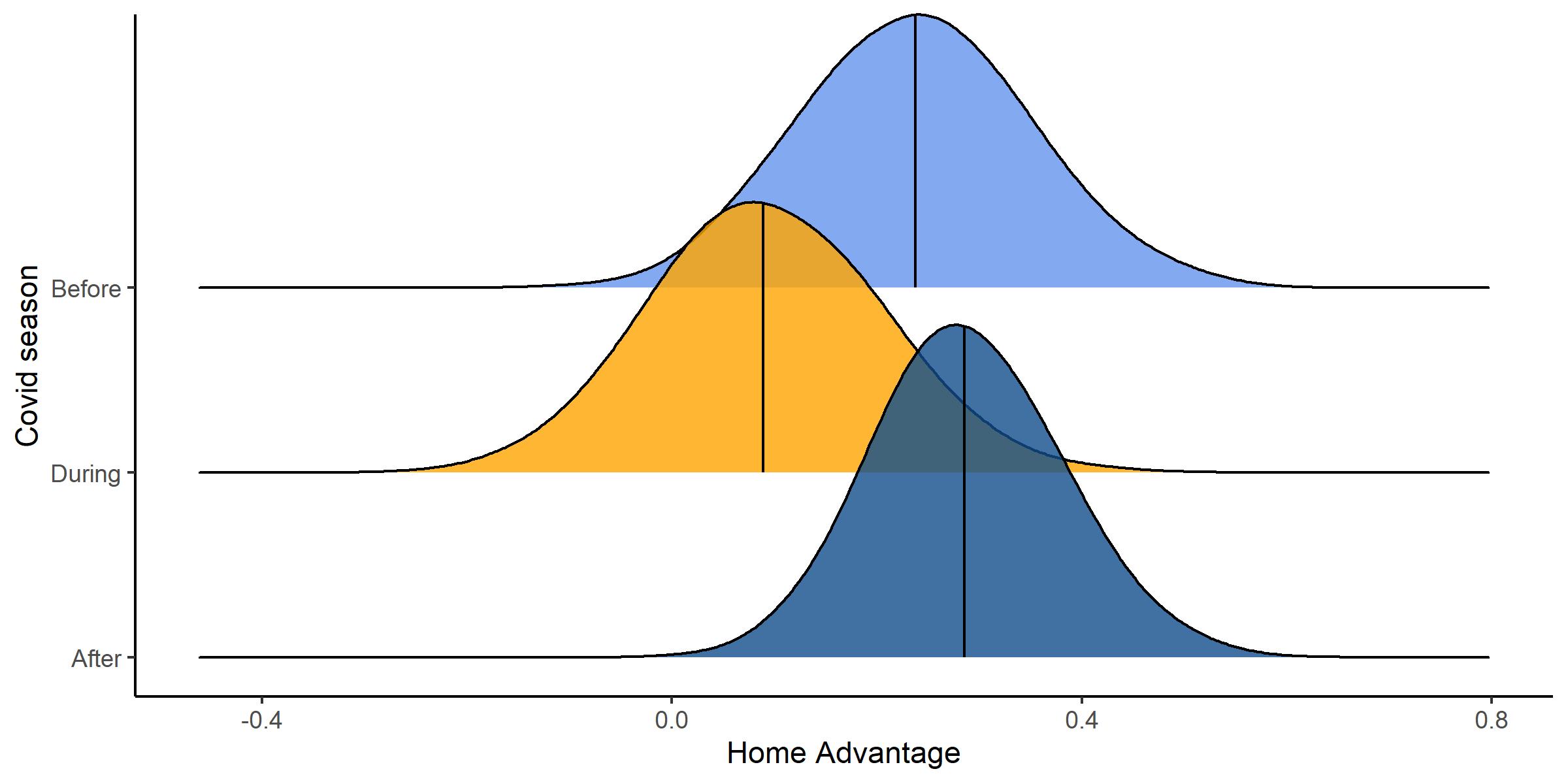}
    \caption{Analysis of Premier League data: distribution of the HA for pre, during, and post-COVID season in the Premier League}
    \label{soccer_covid}
\end{figure}

In Figure \ref{soccer_covid}, we examine the distribution of the HA before the COVID-19 pandemic ($HA_{B}$), during the pandemic ($HA_{D}$), and after the pandemic restrictions ($HA_{A}$). We define the HA effect in the log scale as the rate of the average home team goals and the away goals during the corresponding COVID-19 stage, i.e, $HA_{B} = log(\frac{\mu_1}{\mu_2}) = \beta_{H} + c_H - (\beta_{A} + c_A), HA_{D} = \beta_{H} - \beta_{A}$, and $HA_{A} = \beta_{H} + c'_H - (\beta_{A} + c'_A)$. The estimated effects of $HA_{B}$, $HA_{D}$, and $HA_{A}$ are approximately 0.238, 0.0916, and 0.288, respectively. This suggests that we initially expected the Home Team to score 26.8\% more goals than the Away Team before the pandemic. However, during the pandemic, this advantage reduced to 9.6\%, and after the COVID-19 restrictions, it appeared to recover to 33.4\%.  Furthermore, we find that the HA was significantly diminished during the COVID-19 pandemic, as evidenced by the posterior probability $P(HA_{D} < HA_{B}| \text{data}) = 0.8525$. However, after the COVID-19 restrictions were lifted, there seems to be a recovery in the HA, as indicated by the  posterior probability $P(HA_{D} < HA_{A}|\text{data}) = 0.9527$. Among the reasons as to why the HA was negatively affected during the COVID-19 pandemic and subsequently recovered as restrictions were lifted, some authors have provided evidence suggesting that the absence of crowds influenced the behavior of referees in soccer, resulting in fewer penalties being awarded to away teams \citep{mccarrick2020home, tilp2020covid}. In recent work by \cite{price2022much}, they evaluated the HA in soccer using a causal inference approach. They found that the HA is primarily driven by offensive-based statistics, where the home team gains an advantage by dominating the field, resulting in more shots, passes, and opportunities in the danger zone. Conversely, other authors suggest that crowd influence plays a significant role in the HA effect \citep{pettersson2010behavior, lopez2016persuaded, reade2020echoes}. These perspectives highlight the multifaceted nature of the HA effect and the various factors that can contribute to its manifestation in different sports.

\begin{figure}[!ht]
    \centering
    \includegraphics[scale=0.5]{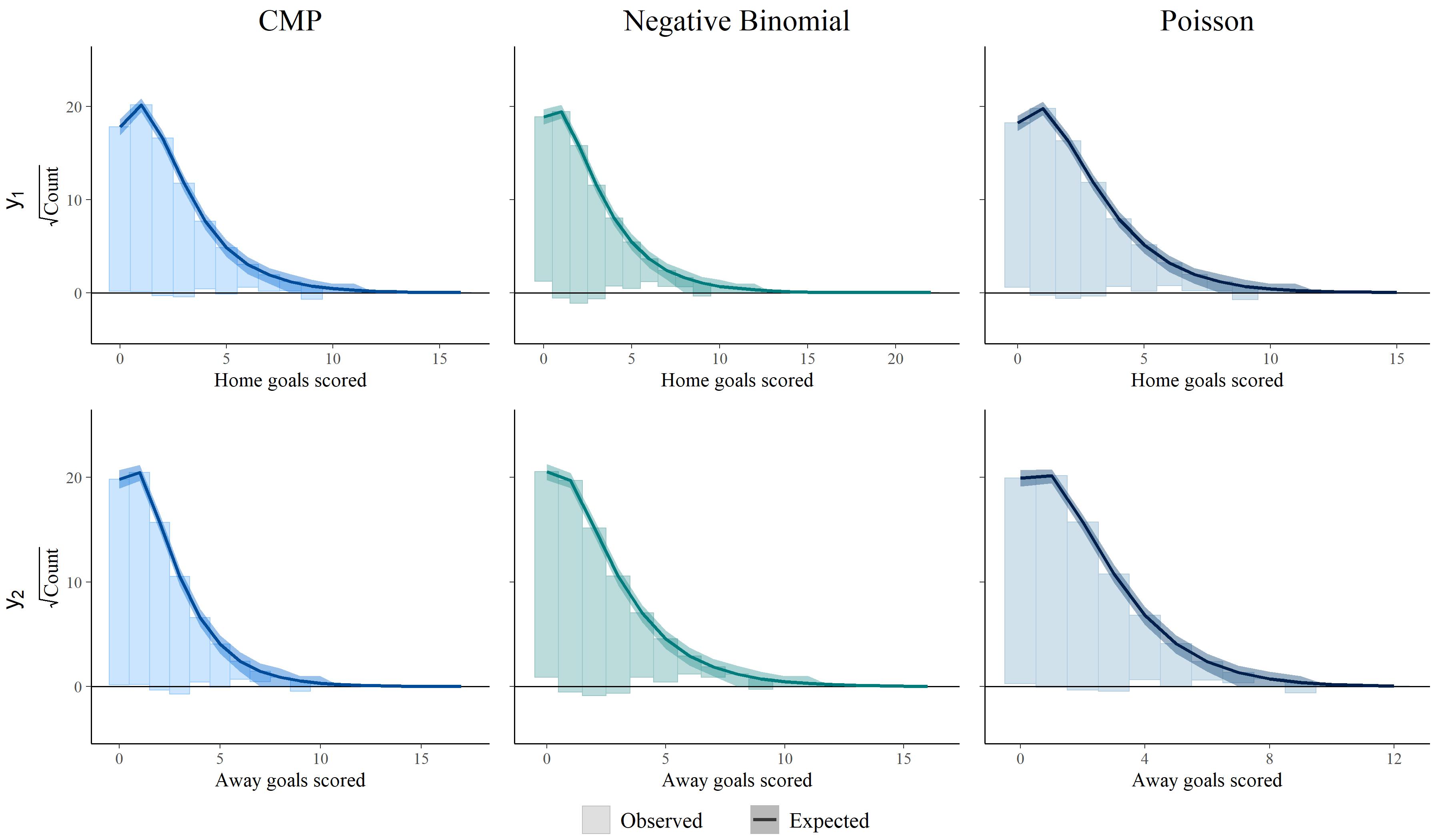}
    \caption{Analysis of Premier League data: posterior predictive Rootogram of home team goals $y_1$ and away team goals $y_2$ for the CMP, Negative Binomial and Poisson models}
    \label{fig:compare_soccer}
\end{figure}

Finally, in Figure \ref{fig:compare_soccer}, we compare the fitting of our proposed model with the Negative Binomial and Poisson models. The posterior predictive rootograms \citep{kleiber2016visualizing} for the home and away team goals are displayed for each distribution. The rootograms provide a visual illustration of the goodness of fit, showing the alignment between the observed counts and the expected counts for each distribution. In the plot, the observed counts' bars are shown hanging from the curve that represents the expected counts. A perfect fit occurs when the observed bars coincide with the horizontal line at 0.  
Upon examination, it is evident that both the CMP  and Poisson models exhibit similarly good fits to the data. The CMP distribution appears to effectively capture the pattern of the observed home team goals, especially for zero and lower counts. {\color{black} Additionally, in Table \ref{tab:dic_soccer} we  display the deviance information criterion (DIC) \citep{spiegelhalterbayesian}. A smaller DIC indicates a better fit to the data set. As it can be seen, the lowest values correspond to the proposed CMP model.}

\begin{table}[!ht]
    \centering
    \caption{Analysis of Premier League data: Deviance Information Criterion (DIC) for the CMP, Negative Binomial (NB) and Poisson models. Bolded numbers represent the lowest DIC for each response}
    \begin{tabular}{lrr}
    Model & $y_1$ & $y_2$ \\ \midrule
    CMP   & \textbf{3635.611}  &  \textbf{3475.992} \\   NB   & 3759.975  &  3562.146 \\
    Poisson & 3738.665 & 3543.986 \\
    \end{tabular}
    \label{tab:dic_soccer}
\end{table}

\section{Analysis of Major League of Baseball data}
\label{data_analysis_baseball}
\subsection{Data description}

\begin{figure}[!ht]
    \centering
    \includegraphics[scale = 0.5]{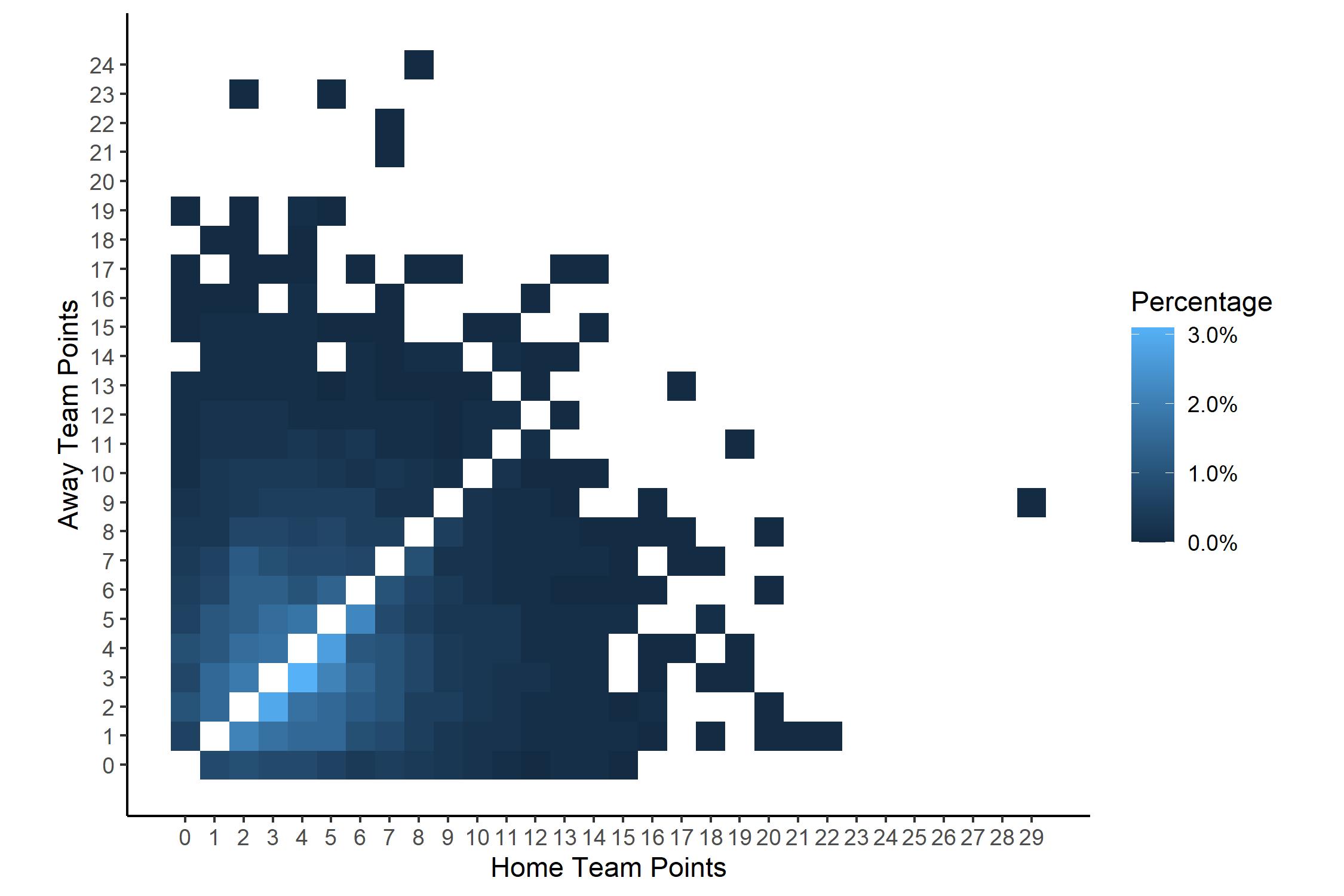}
    \caption{Analysis of MLB data: distribution of scores in the MLB data}
    \label{fig:baseball_runs}
\end{figure}

To further elucidate the performance of the CMP model on an additional dataset, we conducted an analysis of 5,756 Major League Baseball (MLB) games played during the regular seasons of 2019, 2020, and 2021.\footnote{Data extracted from: \url{https://www.retrosheet.org/}} The average number of runs scored by the home teams was 4.72 per game, with a variance of 10.24. Conversely, the away teams averaged 4.63 runs per game, with a variance of 11.01. Notably, both sets of data exhibit a dispersion statistic, $\sigma_p$, greater than 2.1, indicating the data is significantly over-dispersed \cite{payne2018empirical}. In contrast to the soccer data, baseball scores appear to exhibit no correlation, as evidenced by the Spearman correlation coefficient of 0.004. The relative frequency of the home team scoring more runs than the away team, denoted as $Fr($Home Team runs $>$ Away Team runs$)$, is 0.534, while the frequency of the home team scoring fewer runs or an equal number of runs, denoted as $Fr($Home Team runs $\leq$ Away Team runs$)$, is 0.46. These statistics suggest a general positive HA effect. The distribution is visualized in Figure \ref{fig:baseball_runs}. It is important to note that in baseball, ties do not occur. If a game is tied after nine innings, extra innings are played until one team scores and the other does not.

\begin{figure}[!ht]
   \centering
    \includegraphics[scale = 0.06]{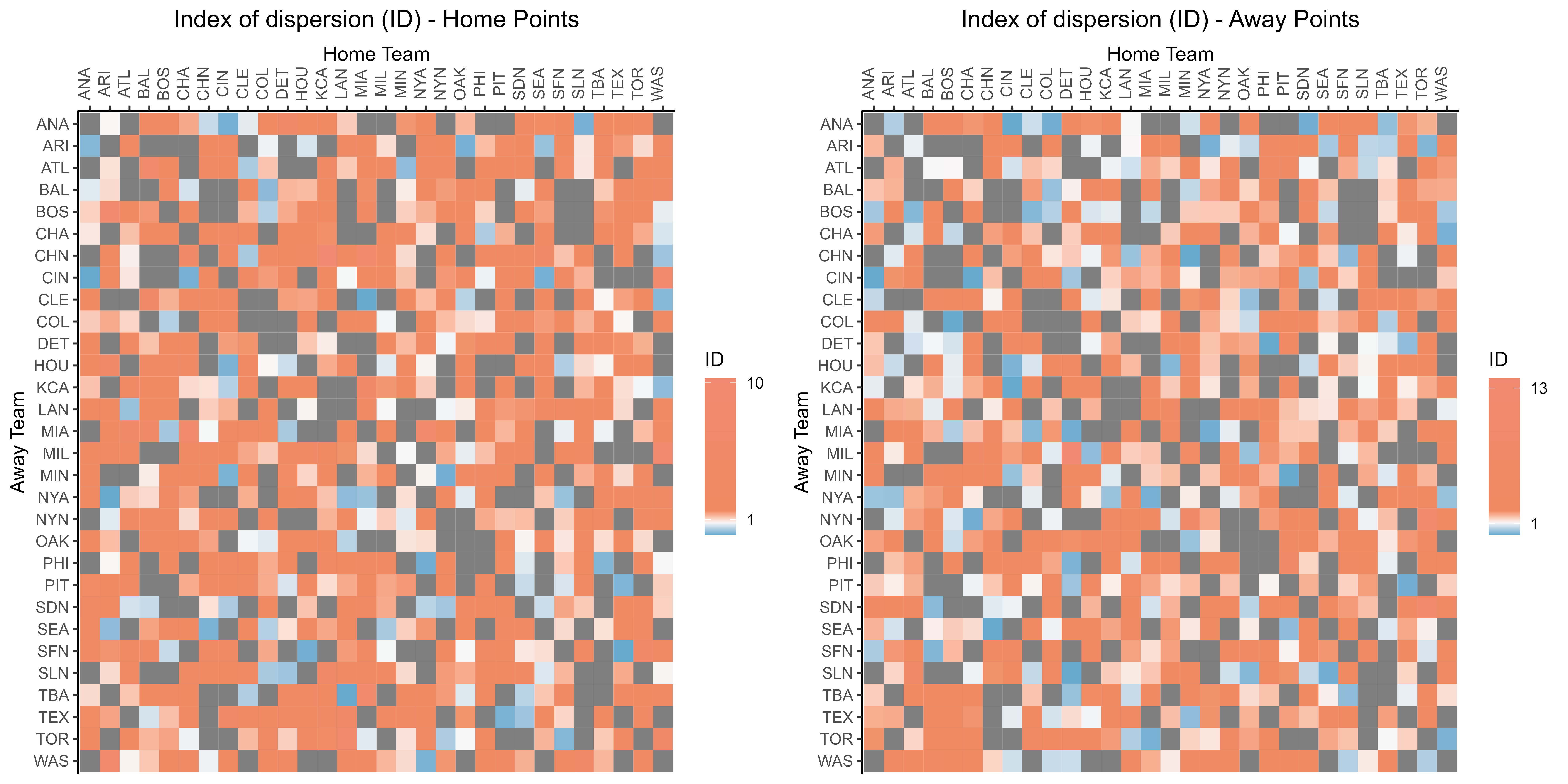}
    \caption{Analysis of MLB data: Indices of dispersion (ID)}
    \label{fig:ID_mlb}
\end{figure}

In  Figure \ref{fig:ID_mlb} we find the Index of dispersion (ID) for the MLB data. In this case, we observe a natural tendency for data to be over-dispersed, especially in the number of points scored by the home team. Furthermore, the index of dispersion varies for specific teams. For instance, when playing at home, the ID of the number of runs of the Toronto Blue Jays (TOR) is greater than 1 regardless of the team they are playing against (TOR column, left heatmap). However, the number of runs of some opponents (ARI, SEA, SFN, WAS) could be slightly equi- or under-dispersed (TOR column, right heatmap). These analyses support the proposed model, as we observe significant variation in the dispersion of goals or points among different teams, and suggest that the effect differs when playing at home or away.

\begin{figure}[!ht]
	\centering
	\subfigure[Average runs per game playing at home field]{\includegraphics[width=0.45\textwidth]{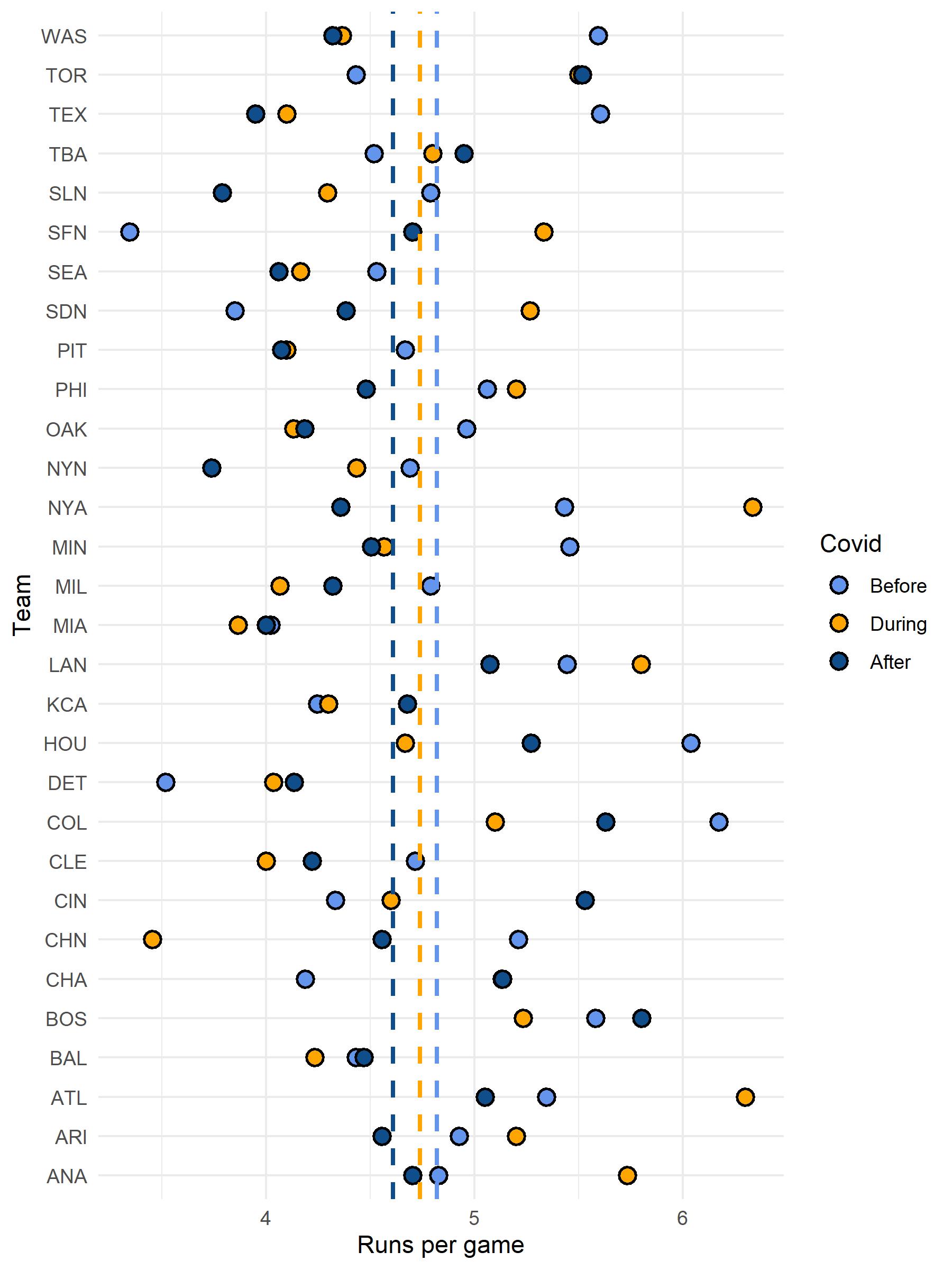}
		\label{fig:avg_home_mlb}}
	\hspace{0.05\textwidth} % Add horizontal space between the subfigures
	\subfigure[Average runs per game playing at away field]{\includegraphics[width=0.45\textwidth]{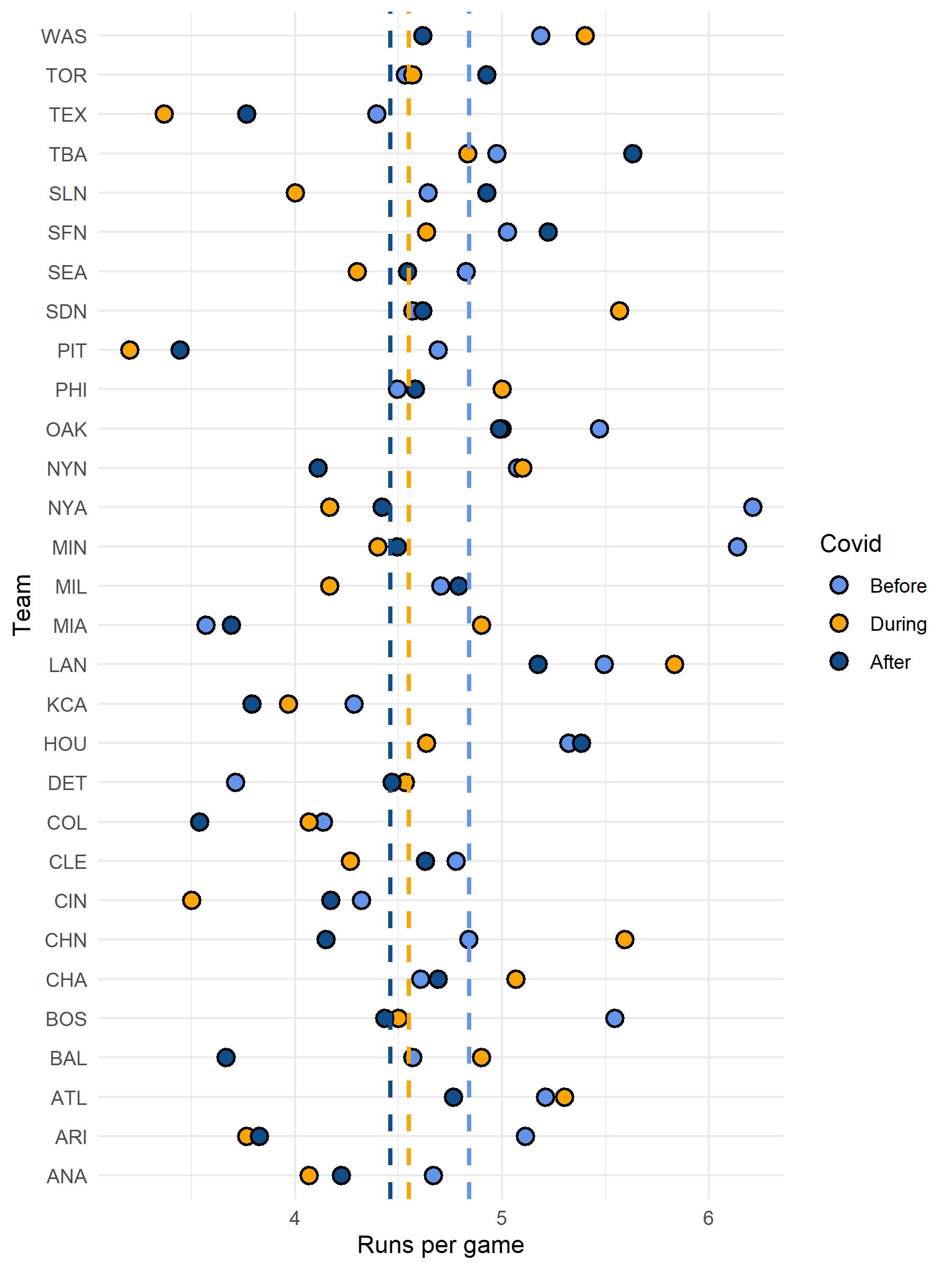}
		\label{fig:avg_away_mlb}}
	\caption{Analysis of MLB data: average team runs per game by Covid season}
	\label{fig:avg_mlb}
\end{figure}

Baseball, in comparison to other sports, is known for having a smaller HA \citep{jones2015home}. In the dataset analyzed, we found that the average number of runs scored by the home team before the pandemic, during the pandemic, and after the pandemic were 4.82, 4.74, and 4.61, respectively. Conversely, the average number of runs scored by the away team before the pandemic, during the pandemic, and after the pandemic were 4.84, 4.55, and 4.46, respectively. Figure \ref{fig:avg_mlb} provides a visual representation of the average runs per game for all teams during different stages of the COVID-19 pandemic.  Comparing the pandemic period to the 2019 season, we observe that 17 out of the 30 teams decreased the number of runs scored when playing at home, while 15 teams decreased the number of runs scored when playing away. However, the decrease was more pronounced in the average number of runs scored by away teams, suggesting a potential increase in the home advantage.

%Furthermore, Figure \ref{fig:dist_goals_covid} presents the distribution of home point differences across seasons. A positive difference indicates a win for the home team, while a negative difference indicates a win for the away team. We noticed a different trend in the 2020 season, with the proportion of home team wins increasing to 55.12\% compared to 52.94\% in the 2019 season, before decreasing again to 53.8\% in 2021. To assess if there were any significant differences in the mean home point differences across seasons, we performed an ANOVA test. The resulting p-value of 0.357 suggests that no significant differences were detected by the test.

Additionally, in Figure \ref{fig:dist_goals_covid}, we displayed the distribution of the home difference of runs across seasons. A positive difference indicates a win for the home time, while a negative difference indicates a win for the away team. With this, we observed a different trend in the 2020 Season, with the proportion of home team wins increasing to $55.12\%$ compared to $52.94\%$ in the 2019 season, before decreasing again to $53.8\%$ in 2021. Similar to the soccer case, an ANOVA test was performed trying to detect any significant difference in the mean of home points differences across seasons. The resulting p-value of 0.357 indicates that there are no significant differences detected by the test.

\begin{figure}[!ht]
    \centering
    \includegraphics[scale = 0.6]{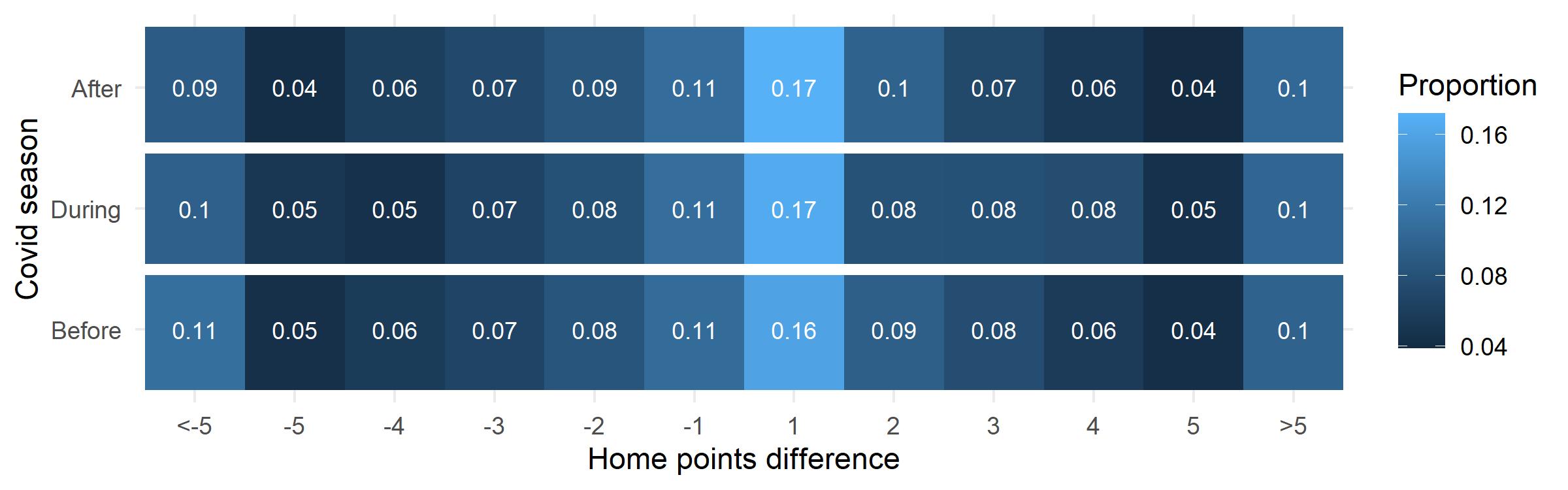}
    \caption{Analysis of MLB data:  distribution of the differences between home team runs and away team runs}
    \label{fig:dist_goals_covid}
\end{figure}

\subsection{Model fitting}

\begin{figure}[!ht]
	\centering
	\includegraphics[scale = 0.6]{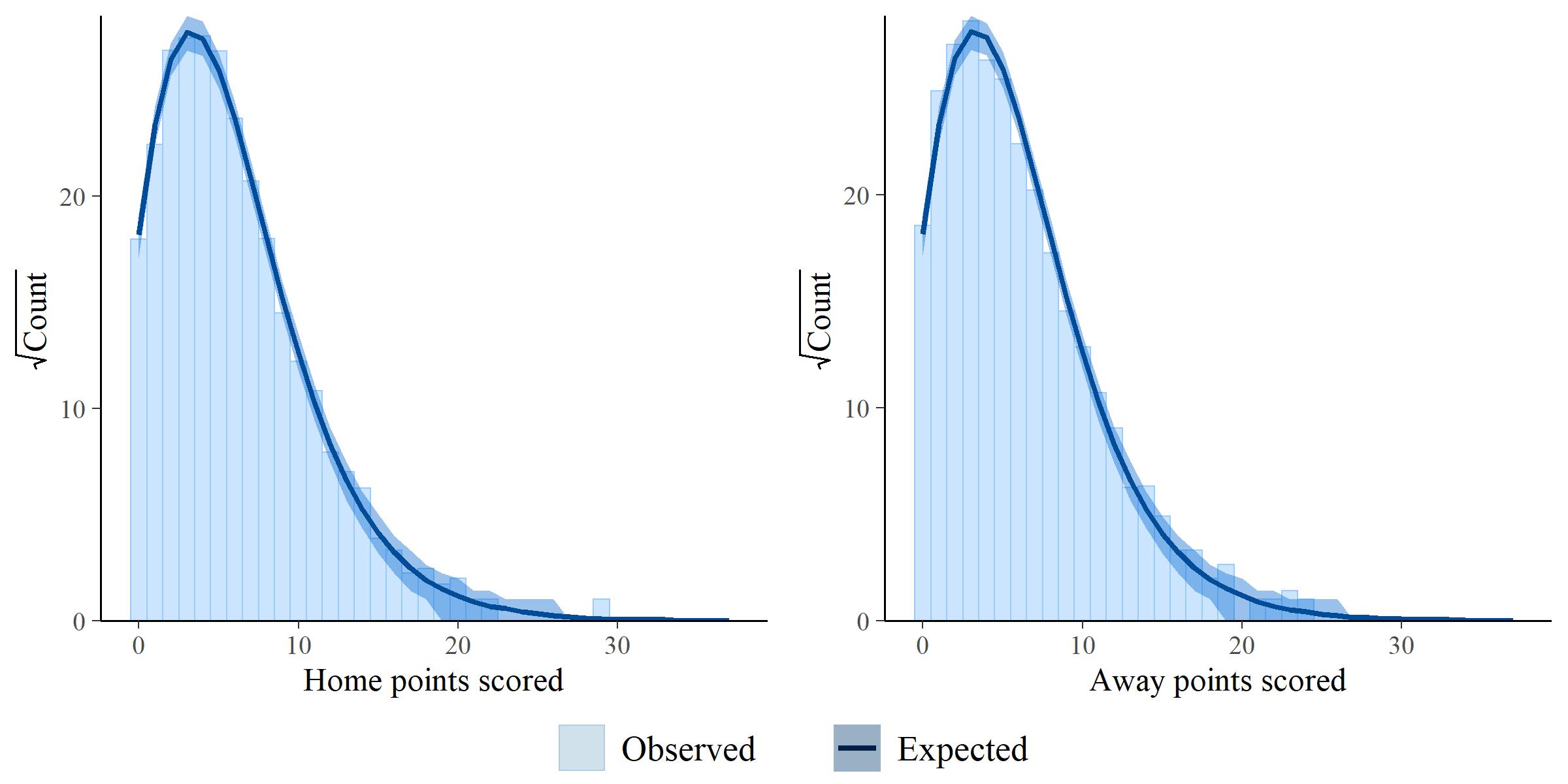}
	\caption{Analysis of MLB data: empirical distribution and the posterior predictive distribution of the home goals and away points in MLB data}
	\label{fig:baseball_fit}
\end{figure}

Although the data reveals no apparent correlation between the scores of the home team and the away team, it does exhibit significant over-dispersion that can be effectively captured by the proposed CMP model.   To fit the model, we assumed again weak prior information, setting $\beta_0 = \gamma_0 = 0$, $B_0 = G_0 = 0.1I$, $\nu_0 = 50$, and $R_0 = I$. {\color{black} We used 180K MCMC iterations with a burn-in of 50K iterations, which resulted in an effective sample size greater than 400 for all the parameters}. The resulting posterior predictive distribution is displayed in Figure \ref{fig:baseball_fit}, illustrating a well-fitting distribution that closely aligns with the observed data.

\begin{figure}[!ht]
    \centering
    \includegraphics[scale = 0.4]{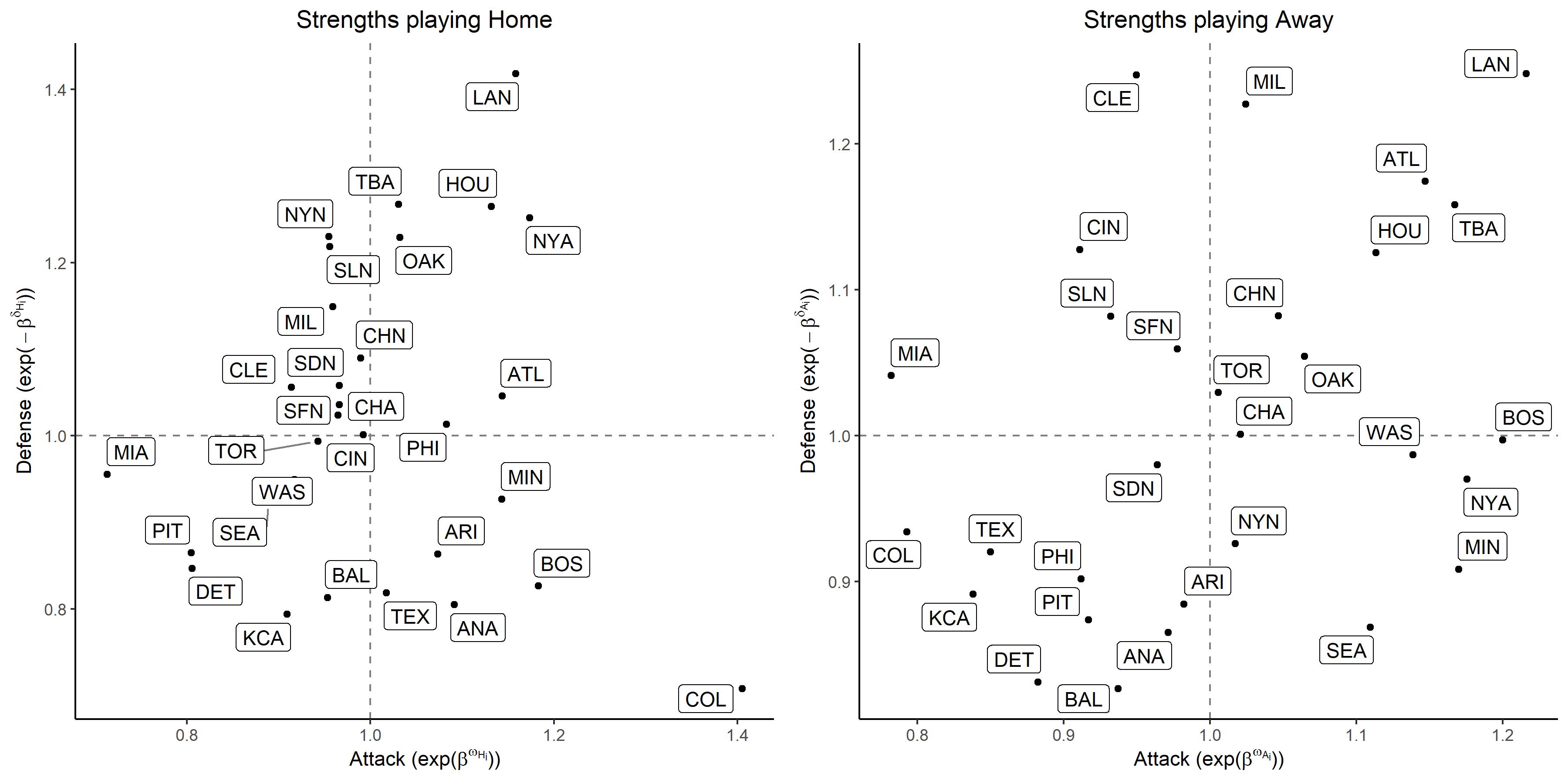}
    \caption{Analysis of MLB data: attack and defense strength of each team}
    \label{fig:posterior_mlb}
\end{figure}

In Figure \ref{fig:posterior_mlb}, we can observe the strength of defense and attack for each team when playing on their home field and when playing away. Notably, the Los Angeles Dodgers (LAN), Atlanta Braves (ATL), and Houston Astros (HOU), winners of the last three World Series Championships, are among the teams with strong performance in both home and away games. On the other hand, the Colorado Rockies (COL) exhibit a contrasting pattern. While they have one of the weakest attacks when playing away, they possess the strongest attack when playing at home. This can be attributed to the unique circumstances of their home stadium, Coors Field, which is situated at an elevation of 5,183 feet above sea level. The Colorado Rockies may benefit from their familiarity and adaptation to playing under these conditions, which could account for their notable advantage when competing at home. {\color{black} An opposite behavior is observed with the Seattle Mariners (SEA), as their strength in attacking is better on the road than at home. During 2022, they also scored more runs away than at home, emerging as one of MLB's top road teams with 41 wins. Manager Scott Servais suggested that some road ballparks are more conducive to home runs than their home ballpark (T-Mobile Park) \citep{Kramer2022-lr}. Estimation of the shape parameters are reported in the Appendix.}

\begin{figure}[!ht]
	\centering
	\includegraphics[scale = 0.5]{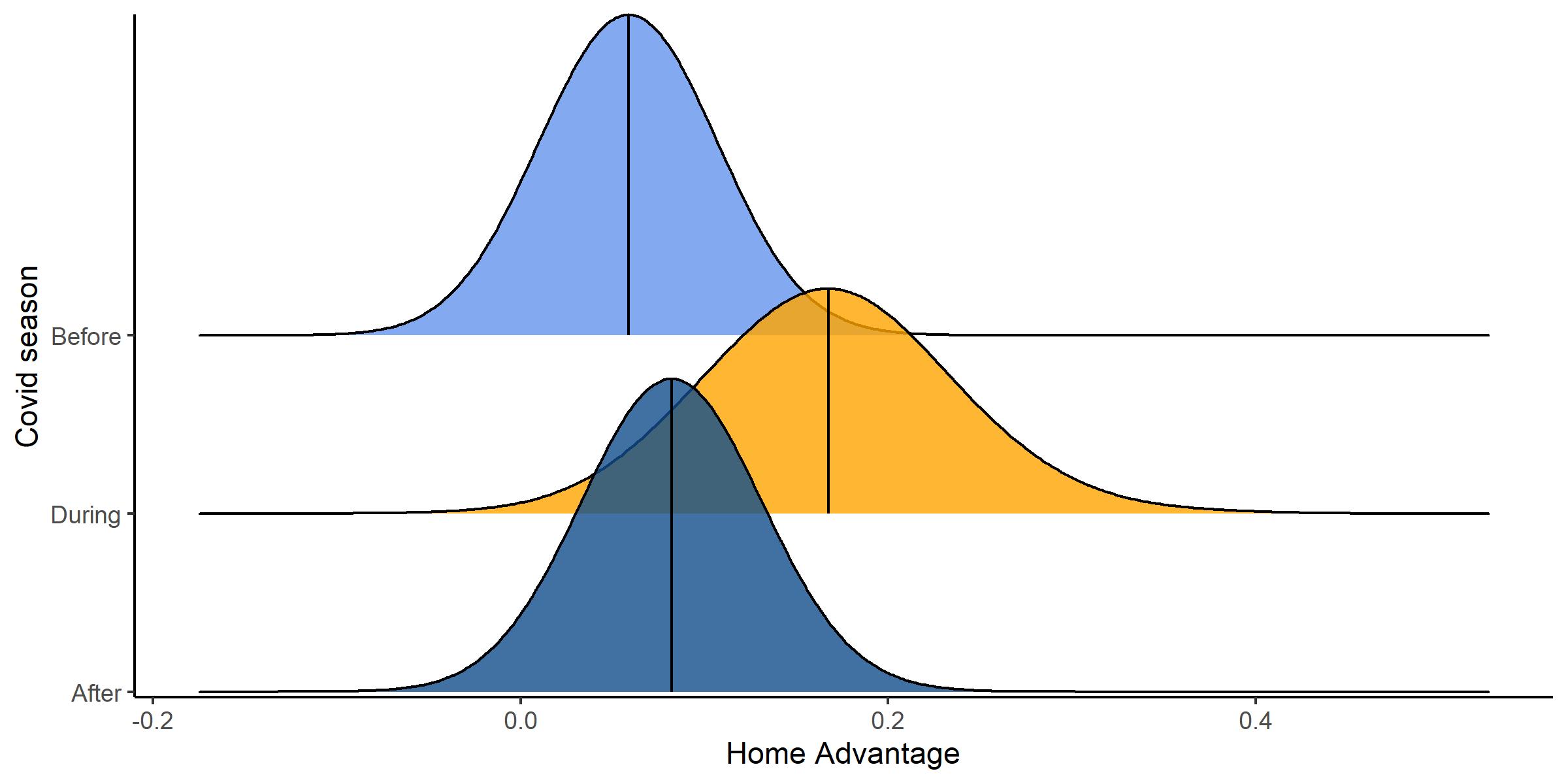}
	\caption{Analysis of MLB data: distribution of the HA for pre, during, and post-COVID-19 season in the MLB}
	\label{fig:cov_baseball}
\end{figure}

As depicted in Figure \ref{fig:cov_baseball}, the HA effect demonstrated an increase during the 2020 season compared to both the 2019 and 2021 seasons. 
Notably, studies such as \cite{higgs2021bayesian} and \cite{chiu2022major} suggest that, despite a slight increase in the HA effect in baseball, it was not significantly affected during the COVID-19 pandemic. It is important to acknowledge a limitation of this analysis, namely that we only considered one season before the pandemic; thus, the observed difference during the COVID-19 season may not be statistically significant when considering previous years. However, these findings lend support to the hypothesis that the HA effect in baseball behaves differently compared to other sports, where it was negatively impacted by the pandemic \citep{higgs2021bayesian, mccarrick2020home}.

Our observations align with the prevailing notion that the HA in baseball tends to be smaller compared to soccer. In soccer, the HA effect is often more pronounced and can have a significant impact on game outcomes. 
%Several theories have been put forth to explain the HA observed in baseball. One theory suggests that the HA is not primarily associated with the presence of spectators but rather stems from factors such as facility familiarity or travel fatigue \cite{chiu2022major}. This implies that players may perform better when they are accustomed to the specific characteristics of their home field or when they are less fatigued from travel. Another perspective proposes that the HA in baseball arises from the unique nature of the sport, which is often considered more individualistic compared to other team sports. In baseball, the actions of individual players, such as pitchers and batters, have a substantial impact on the outcome of the game. Therefore, the team dynamics and the boost from home crowd energy that players in other sports may experience are not as prominent in baseball \cite{jones2015home, losak2021baseball}.
This difference can be attributed to various factors, including cultural differences and the behavior of fans in stadiums. In baseball, the HA effect is thought to be more associated with factors such as familiarity with the home field or travel fatigue, as suggested by \cite{chiu2022major}. Players may have a better understanding of the unique characteristics of their home field, providing them with a slight advantage. Additionally, the physical demands of travel in baseball, with teams frequently playing on the road, may contribute to a smaller HA effect. Another perspective suggests that the dynamics of baseball as a sport contribute to a smaller HA effect compared to team sports like soccer. Baseball is often considered more individualistic, with the actions of individual players, such as pitchers and batters, playing a significant role in the outcome of the game. This individual nature of the sport may reduce the impact of the HA effect, which is more pronounced in team sports where the collective dynamics and energy from the crowd can influence player performance \citep{jones2015home, losak2021baseball}. These theories highlight the complex interplay of factors contributing to the HA effect in baseball, which may differ from those observed in other sports.

\begin{figure}[!ht]
    \centering
    \includegraphics[scale = 0.5]{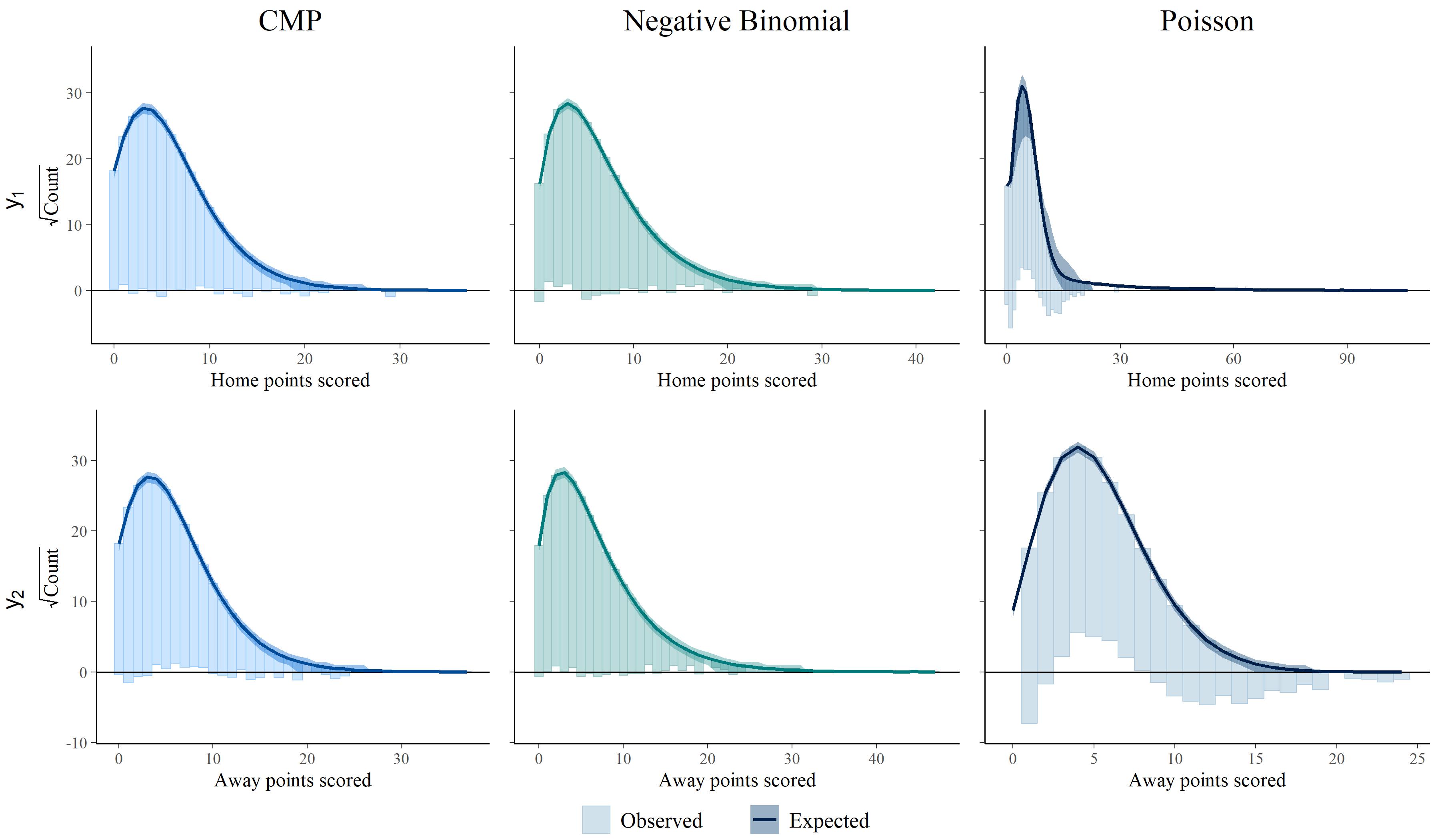}
    \caption{Analysis of MLB data: posterior predictive rootogram of home team points $y_1$ and away team points $y_2$ for the CMP, Negative Binomial and Poisson models}
    \label{fig:comparison_mlb}
\end{figure}

In a similar manner to the analysis of soccer data, we conducted a visual comparison of the fitting performance of our proposed CMP model with that of the Negative Binomial and Poisson models using rootograms, as shown in Figure \ref{fig:comparison_mlb}. The rootograms provide a visual representation of the deviations between the observed data and the expected values from each model.
Upon inspection, it becomes apparent that the Poisson model is not suitable for accurately modeling the baseball data due to the evident over-dispersion present in the dataset. Conversely, the CMP model demonstrates a strong performance and is a compelling alternative to the widely used Negative Binomial model in this context. The rootograms exhibit fewer and smaller deviations for the CMP model, indicating a better fit to the considered datasets. {\color{black} To complement this analysis, we calculated the Deviance Information Criterion (DIC) for all the models. Results are presented in Table \ref{tab:dic_mlb}.} This underscores the robustness of the proposed CMP model in accommodating various types of data dispersion and its ability to achieve a favorable fitting outcome.

\begin{table}[!ht]
	\centering
	\caption{Analysis of MLB data: Deviance Information Criterion (DIC) for the CMP, Negative Binomial (NB) and Poisson models. Bolded numbers represent the lowest DIC for each response}
	\begin{tabular}{lrr}
		Model & $y_1$ & $y_2$ \\ \midrule
		CMP   & \textbf{27943.03}  &  \textbf{28135.32} \\       
		NB   & 28516.47  &  28564.52 \\
            Poisson & 65994.64 & 30906.11 \\
	\end{tabular}
	\label{tab:dic_mlb}
\end{table}

\section{Discussion and conclusion}
\label{conclusion}

In this paper, we have demonstrated the practical application of a bivariate CMP model for analyzing count data in sports, with a specific focus on investigating the HA effect in soccer and baseball, while also considering its variations during the COVID-19 pandemic. The flexibility of the proposed model allows to effectively handle count data exhibiting different levels of dispersion, including under-dispersion, over-dispersion, and equi-dispersion. To account for the potential correlation between home and away scores, we have incorporated random effects into the model.
%, as first suggested by Chib et al. (2001).  
We have demonstrated the utilization of the rejection sampler proposed by \cite{benson2021} and the Exchange algorithm presented by \cite{murray2012} to perform inference. These methods offer computational efficiency in the estimation of the model parameters.
%{\color{blue}, however, these algorithms are more complex than simpler models, and they require more iterations to reach convergence.} 

Results of our analyses have indicated that the proposed CMP model is robust and capable of effectively modeling count data with any level of dispersion. Indeed, data applications and simulations have shown that our model is {\color{black} not only }capable of adapting to different types of data, consistently matching or outperforming the fitting of standard Negative Binomial and Poisson models{\color{black}, but also reliable in the estimation of the true parameters regardless of the dispersion of the data. It is worth noting, however, that the proposed model increases the computational time per iteration with respect to fitting simpler models. For example, under the one season scenario, 1000 iterations of our model require approximately 30 seconds, while other models take about 15 seconds. Moreover, due to the implementation of the Exchange Algorithm, a larger number of ierations is required to achieve a given effective sample size.
We have demonstrated that when modeling under or over-dispersed data, estimations from the standard models can be biased due to the dispersion of the data in contrast to the proposed CMP model.} This highlights the reliability and versatility of our model when modeling data from various leagues, seasons, or sports, where the level of dispersion may vary. 
%It provides a robust alternative to ensure a good fit in diverse scenarios. 

In our analyses of the soccer and baseball data, we found that the HA in soccer was negatively affected during the COVID-19 pandemic but subsequently recovered as restrictions were lifted. In contrast, we found that the HA in baseball, in general, is smaller than in soccer, supporting the various theories that have been put forth to explain this phenomenon, including cultural differences and the behavior of fans in stadiums. It is worth noting that in baseball,  if the home team is winning after eight and a half innings, then the home team is excluded from batting in the bottom half of the ninth inning. This particular rule could slightly deflate the estimated HA in baseball. It is also important to note that the model's performance can be sensitive to the choice of random effects, particularly the variance components. Therefore, careful consideration should be given to selecting appropriate priors that reflect prior knowledge or assumptions about the variability of the random effects. 
%This difference can be attributed to various factors, including cultural differences and the behavior of fans in stadiums. In baseball, the HA effect is thought to be more associated with factors such as familiarity with the home field or travel fatigue, as suggested by \cite{chiu2022major}. Players may have a better understanding of the unique characteristics of their home field, providing them with a slight advantage. Additionally, the physical demands of travel in baseball, with teams frequently playing on the road, may contribute to a smaller HA effect. Another perspective suggests that the dynamics of baseball as a sport contribute to a smaller HA effect compared to team sports like soccer. Baseball is often considered more individualistic, with the actions of individual players, such as pitchers and batters, playing a significant role in the outcome of the game. This individual nature of the sport may reduce the impact of the HA effect, which is more pronounced in team sports where the collective dynamics and energy from the crowd can influence player performance \cite{jones2015home, losak2021baseball}. 

In summary,  our model offers a flexible framework that can be applied to various settings in sport and beyond sports, enabling the modeling of different types of count data and the potential inclusion of more than two response variables. The code for implementing the model is available at the following GitHub repository: \url{https://github.com/mauroflorez/cmp} (upon acceptance).

%Finally, in both data applications and simulations, we proved that our model is able to adapt to different types of data and even outperforms the fitting of standard Negative Binomial and Poisson models. In this sense, when modeling data from different leagues, seasons, or sports, where the level dispersion can vary, our model provides a reliable alternative to ensure a good fitting in any case.

%Moreover, our model offers a flexible modeling framework that can be useful in sports and other settings, modeling different types of count data, and also it can be adapted to include more than two response variables. The code is available at \url{https://github.com/mauroflorez/cmp}

%Recently, \cite{price2022much} evaluated the HA in soccer using a causal inference approach; they found that the HA resides in offensive-based statistics, where the home team takes advantage by dominating the field, which allows them to have more shots, passes, and reach the danger zone more times. On the other hand, other authors suggest that crowds are an important factor that influences the HA effect \cite{pettersson2010behavior}, \cite{lopez2016persuaded}, \cite{reade2020echoes}. 

%\printbibliography
%\bibliographystyle{apalike}
%\bibliography{myref}

\newpage 

\appendix

\section*{Appendix}

\subsection*{A. Proof of Equation \eqref{proof}}

\begin{proof}
Let $\delta_i = (\delta_{i1}, \delta_{i2})$, where $\delta_{ij} = \exp(b_{ij})$, then $\delta_i = \exp(b_i) \sim LN_2(\mu, \Sigma)$ with $\mu = \exp(0.5\cdot diag(D))$, and $\Sigma = diag(\mu)(\exp(D) - \mathbf{11^{T}})diag(\mu)$. Also, assume that $\lambda_{ij} = \exp{(x_{ij}^T\beta_j)}$, then $y_{ij}|\lambda_{ij}, \delta_{ij}, \nu_{ij} \sim CMP(\lambda_{ij}\delta_{ij}, \nu_{ij})$. Finally, let $\widetilde{\lambda}_{ij} = \lambda_{ij}\mu$, $\widetilde{\lambda}_i = (\widetilde{\lambda}_{i1}, \widetilde{\lambda}_{i2})$, and $\widetilde{\Lambda_i} = diag(\widetilde{\lambda}_i)$. By law of iterated expectations and using the asymptotic approximations of \cite{shmueli2005}, we have that $E(y_i|\beta,\gamma, D) \approx \widetilde{\lambda}_{i} + \frac{1}{2\nu_i} - \frac{1}{2}$, and $var(y_i|\beta,D) \approx V_i^{-1}\widetilde{\Lambda}_{i}(\exp(D) - \mathbf{11^{T}})\widetilde{\Lambda}_{i}$, $V_i = diag(\nu_{i1}, \nu_{i2})$.
Thus
\begin{equation*}
    \begin{split}
        cov(y_{i1}, y_{i2}) &\approx \widetilde{\lambda}_{i1}(\exp(d_{12})-1)\widetilde{\delta}_{i2} \\
        &\approx \lambda_{i1}\exp(0.5 d_{11})(\exp(d_{12})-1)\lambda_{i2}\exp(0.5d_{22}).
    \end{split}
\end{equation*}
Clearly, \eqref{proof} can be positive or negative depending on the sign of $d_{12}$, i.e, the non-diagonal element of $D$.
\end{proof}

\subsection*{B. Prior Sensitivity analysis}
\label{sec:simulations_2}

We assess the performance of the bivariate Conway-Maxwell-Poisson (CMP) model under various prior specifications. Specifically, we focus on the Over-dispersed scenario of 1 Season discussed in Section \ref{simulations} and examine different combinations of values for $B_0$, $G_0$, $\nu_0$, and $R_0$. These combinations are summarized in Table \ref{tab:sim1_scenarios}. To monitor convergence, we calculate the multivariate potential scale reduction factor ($\hat{R}$) \cite{brooks1998general} using 3 chains and also check the sampler's efficiency by calculating the effective sample size (ESS). %These results are summarized in the Appendix. 
\begin{table}[!ht]
\centering
\caption{Prior sensitivity: values considered in each scenario to assess the sensitivity of the model's performance to the specification of prior hyperparameters}
\begin{tabular}{lrrrr}
Scenarios & A & B & C & D \\ \midrule
$B_0  $ & $0.1I$ & $I$ & $3I$ & $10I$  \\
$G_0 $  & $0.1I$ & $I$ & $3I$ & $10I$  \\
$\nu_0$ & $30$   & $10$ & $10$ &  $5$ \\
$R_0  $ & $I$    & $I$ & $0.1I$ & $0.1I$    \\ 
\end{tabular}
\label{tab:sim1_scenarios}
\end{table}

%Table \ref{tab:estimations} presents the estimated parameters along with their corresponding $95\%$ highest posterior density (HPD) intervals.  It is evident that scenario A yields the most accurate parameter estimates, while scenarios C and D exhibit the least accuracy. This observation aligns with the higher prior expectation for the precision matrix, denoted as $D^{-1}$ (see also the Appendix). Notably, the HPD intervals for the mean of the random variables, $\hat{b}$, demonstrate a noticeable variation. In scenarios C and D, greater heterogeneity is imposed through the values of $\nu_0$ and $R_0$, allowing for larger random effects. In contrast, scenarios A and B describe minimal  prior variability, leading to a significant reduction in the HPD intervals.

Moving on to Table \ref{tab:mse}, we present the mean squared error (MSE) values for $\mu_{j}$ across all scenarios. Consistent with expectations, scenario A demonstrates the best recovery of the observed values. Again, this outcome can be attributed to the small variance of the priors employed in this scenario. Conversely, scenarios C and D exhibit the poorest performance in terms of recovering the observed values.

\begin{table}[!ht]
	\centering
	\caption{Mean Squared Error (MSE) of $\mu$ - Prior sensitivity analysis}
	\begin{tabular}{lrrrr}
		Scenarios & A & B & C & D \\ \midrule
		$\mu_1$   & 0.71  &  1.01 & 1.19    &  1.2  \\       
		$\mu_2$   & 0.25  &  0.25 & 0.22 &  0.545   \\  
	\end{tabular}
	\label{tab:mse}
\end{table}

\begin{table}[!ht]
	\centering
	\caption{Prior sensitivity analysis: multivariate potential scale factor $(\hat{R})$}
	\begin{tabular}{lrrrrrrrr}
		Scenarios                & \multicolumn{2}{c}{A} & \multicolumn{2}{c}{B} & \multicolumn{2}{c}{C} & \multicolumn{2}{c}{D} \\ \midrule
		$\hat{R}$ & $y_1$  & $y_2$  & $y_1$  & $y_2$   & $y_1$  & $y_2$ & $y_1$  & $y_2$     \\
		$\beta$     & 1.07 & 1.06 & 1.15 & 1.16 & 1.2 & 1.21 & 4.22 & 1.82    \\
		$\gamma$    & 1.05 & 1.05 & 1.19 & 1.17 & 1.22 & 1.24 & 4.74 & 5.04		 \\
		$b$         & 1.02 & 1.02 & 1.03 & 1.03 & 1.22 & 1.24 & 4.22 & 3.91		    \\
	\end{tabular}
	\label{tab:sim1_R}
\end{table}

\begin{table}[!ht]
	\centering
	\caption{Prior sensitivity analysis:  Effective Sample Size (ESS)}
	\begin{tabular}{lrrrrrrrr}
		Scenarios                & \multicolumn{2}{c}{A} & \multicolumn{2}{c}{B} & \multicolumn{2}{c}{C} & \multicolumn{2}{c}{D} \\ \midrule
		$\hat{R}$ & $y_1$  & $y_2$  & $y_1$  & $y_2$   & $y_1$  & $y_2$ & $y_1$  & $y_2$     \\
		$\beta$     & 350.44 & 408.54 & 123.25 & 131.13 & 98.48 & 97.13 & 7.26 & 10.71    \\
		$\gamma$    & 506.3 & 511.11 & 195.37 & 192.4 & 106.45 & 111.09 & 47.51 & 43.48 \\
		$b$         & 9366 & 9555.62 & 9175.4 & 9286.9 & 2324.947 & 2351.463 & 2306.58 & 2329.84   \\
	\end{tabular}
	\label{tab:sim1_ESS}
\end{table}

\noindent Scenarios A and B exhibit more stable estimations, with all $\hat{R}$ values below 1.2. Indeed, some of the trace plots for chains in scenarios C and D demonstrated some mild convergence issues (plots not shown).
%, which are further supported by the trace plots provided further below. 
Additionally, the effective sample size (ESS) analysis indicates higher algorithm efficiency in scenario A, while scenarios C and D show lower efficiency. These findings are likely connected to the specified values of $\nu_0$ and $R_0$, as well as the larger covariance matrices $B_0$ and $G_0$. Notably, the chains for random effects display larger $\hat{R}$ values and smaller ESS in scenarios C and D compared to scenarios A and B. This discrepancy may be attributed to the expected value of the inverse of the random effects' covariance matrix $D$, which is determined by $\nu_0 R_0$. Consequently, larger values are permitted for the random effects in scenarios C and D, despite the expectation of low heterogeneity in this simulation scenario.

%Likewise, the ESS shows that the algorithm is more efficient in scenario A and inefficient in scenarios C and D. This is likely to be connected with the values imposed in $\nu_0$ and $R_0$ and the large covariance matrices $B_0$ and $G_0$. We can notice that the chains of the random effects have larger $\hat{R}$ values and smaller ESS in scenarios C and D than in scenarios A and B; this is probably because the expected value of the inverse of the matrix of covariance of the random effects $D$, is given by $\nu_0 R_0$, this allows the random effects to have larger values in scenarios C and D, but we expect to have low heterogeneity in this simulation scenario.  

{\color{black}

\subsection*{C. Shape parameters $\gamma$}

\color{black} In Figure \ref{fig:shape_soccer}, we observe the shape parameters estimation for the Premier League data analysis. In the left sub-figure we have the estimation for parameters associated with the Home Goals. Strong teams, such as Chelsea, Liverpool, and Manchester City, are located below 0, indicating that when other teams play against them, the dispersion parameter will be lower, implying a larger variance in their scored goals (over-dispersion). We observe a similar effect analyzing the Away Goals. Conversely, weaker teams such as Norwich or Brighton have their shape parameter above 0, indicating that when teams play against them, the goals will be more under-dispersed.
On the other hand, in the x-axis, Manchester United and Manchester City are below 0, which means that the number of goals they score tends to be more over-dispersed compared to the other teams. We see the opposite trend with Liverpool, where the goals scored tend to be more under-dispersed. We can compare this with respect to over or under-dispersion because the intercept effect was 0. It is noticeable that the effects are different in the Home Goals and Away scores, supporting the modeling approach we assumed.

\begin{figure}[!ht]
    \centering
    \includegraphics[scale = 0.05]{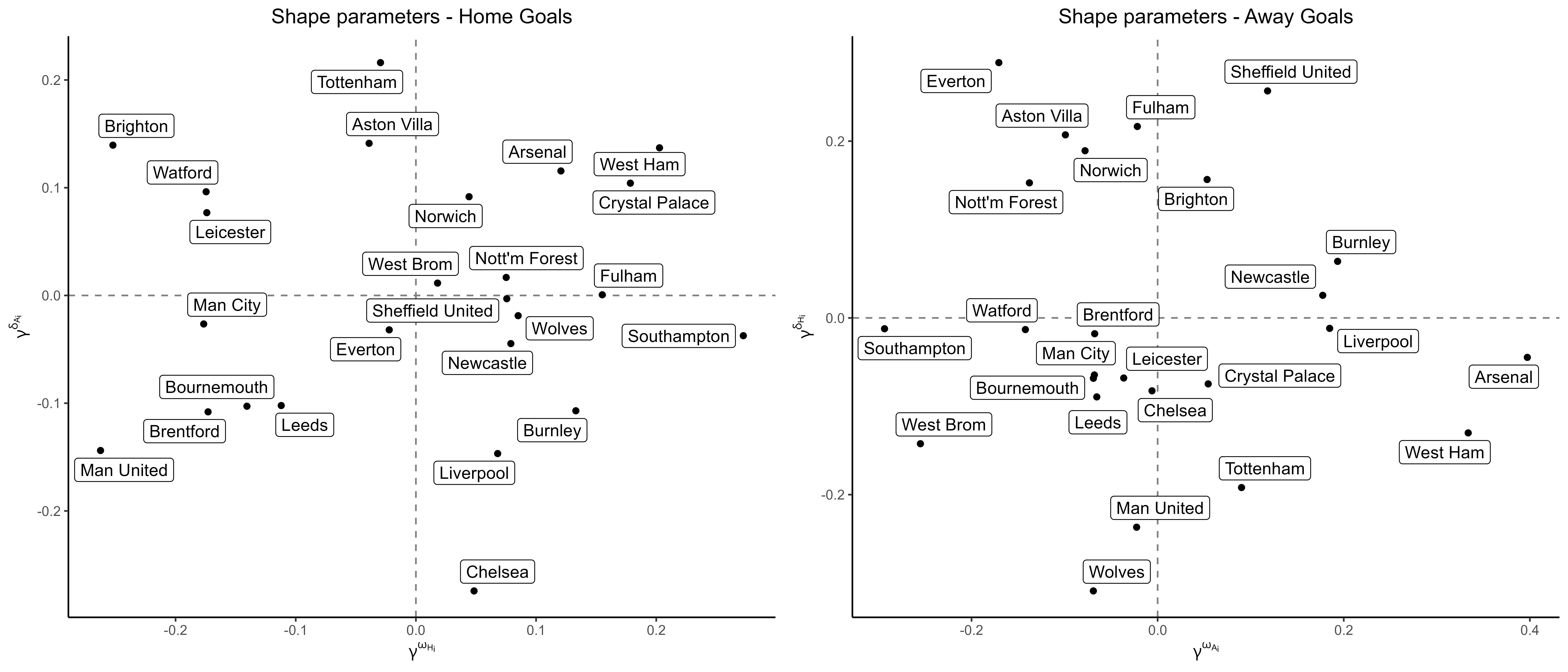}
    \caption{Analysis of Premier League data: Shape parameters}
    \label{fig:shape_soccer}
\end{figure}

Similarly, we plot the estimations for the MLB analysis in Figure \ref{fig:shape_mlb}. One of the noticeable cases is Miami Marlins (MIA), with a negative effect on the x-axis in the Home Points and a positive effect in the Away Points. This means that at Home, MIA tends to have scores more dispersed, while Away, their scores will be less dispersed in comparison to the others. This highlights the ability of our model to adapt to data of different types and to accommodate different phenomena and mechanics observed in the different teams and sports.

\begin{figure}[!ht]
    \centering
    \includegraphics[scale = 0.05]{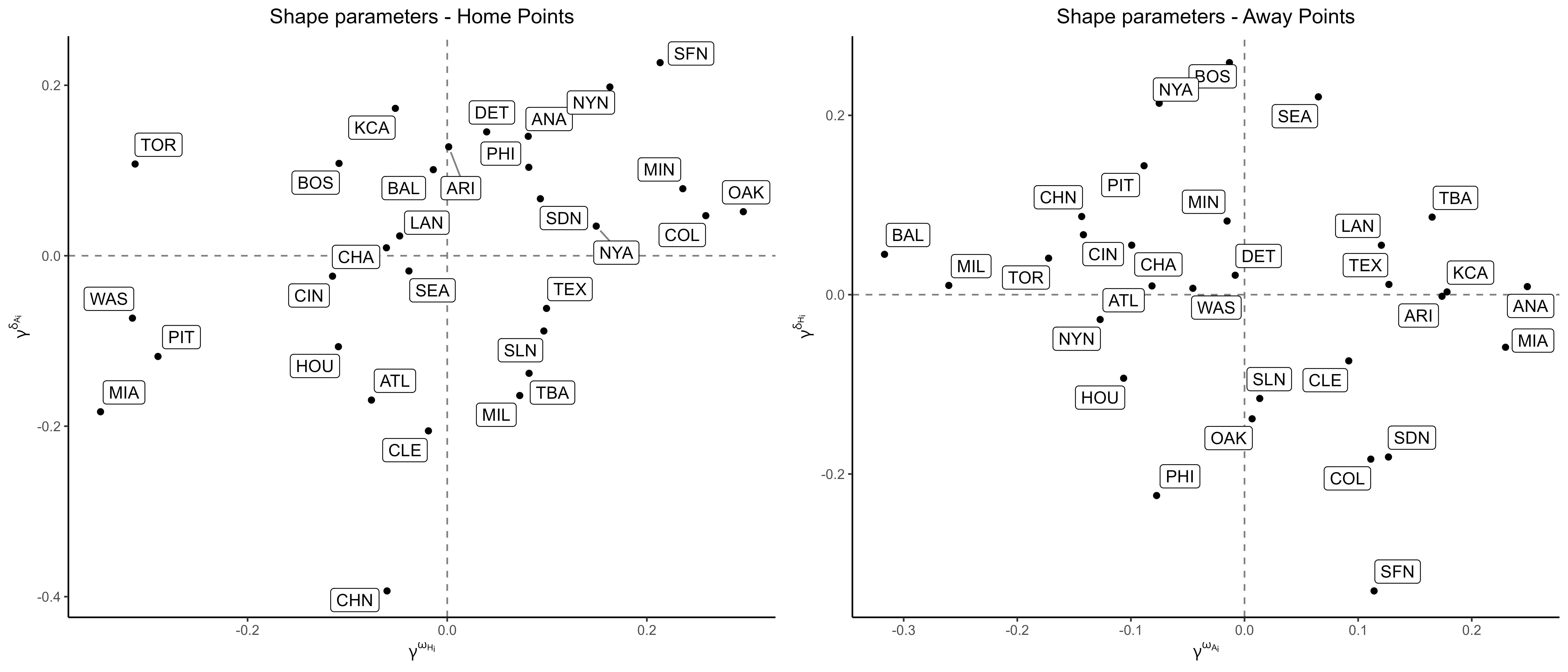}
    \caption{Analysis of MLB data: Shape parameters}
    \label{fig:shape_mlb}
\end{figure}
}

\end{document}